\documentclass[12pt]{article}

\usepackage{amssymb}
\usepackage{amsmath}
\usepackage{latexsym}
\usepackage{yfonts}

\catcode`@=11
\renewcommand{\section}{\@startsection{section}{1}{0pt}{\medskipamount}
{\medskipamount}{\large\bf}}
\numberwithin{equation}{section}
\catcode`@=12

\def\beq{\begin{eqnarray}}    
\def\eeq{\end{eqnarray}}      

\def\pa{\partial}                       

\def\={\ =\ }

\begin{document}

\begin{center}

{\Large\bf Multiplicative renormalization of  Yang-Mills theories in
the background-field formalism}

\vspace{18mm}

{\large Igor A. Batalin$^{(a,b)}\footnote{E-mail:
batalin@lpi.ru}$\;,
Peter M. Lavrov$^{(b, c)} \footnote{E-mail:
lavrov@tspu.edu.ru}$,\;
Igor V. Tyutin$^{(a,b)}\footnote{E-mail:
tyutin@lpi.ru}$\;
}

\vspace{8mm}

\noindent ${{}^{(a)}}$
{\em P.N. Lebedev Physical Institute,\\
Leninsky Prospect \ 53, 119 991 Moscow, Russia}

\noindent  ${{}^{(b)}}
${\em
Tomsk State Pedagogical University,\\
Kievskaya St.\ 60, 634061 Tomsk, Russia}

\noindent  ${{}^{(c)}}
${\em
National Research Tomsk State  University,\\
Lenin Av.\ 36, 634050 Tomsk, Russia}

\vspace{20mm}

\begin{abstract}
\noindent
In the paper, within the background field method, the structure of renormalizations
is studied as for Yang-Mills fields interacting with a multiplet of spinor fields.
By extending the Faddeev-Popov action with extra fields and parameters,
one is allowed to establish the multiplicative character of the renormalizability.
The renormalization of the physical parameters is shown to be gauge-independent.
\end{abstract}

\end{center}

\vfill

\\

\newpage

\section{Introduction}

When quantizing non-Abelian gauge field theories \cite{YM}, whose gauge transformations
form a group,
one is naturally based on the Faddeev-Popov method \cite{FP}.
It is a characteristic property of the Faddeev
-Popov gauge-fixed action that the latter is invariant under global BRST supersymmetry
\cite{BRS1,T},
which, in turn, can be expressed in the form of the Zinn-Justin equation \cite{Z-J}
for the Faddeev
-Popov action. At the quantum level, the BRST symmetry as expressed
in terms of the effective
action, implies the Slavnov-Taylor identities \cite{Tayl,S} to hold.
Further generalization as to the
quantization of gauge theories, including the cases of field-dependent
structure coefficients,
as well as open and/or reducible gauge algebra,
is described by the field-antifield BV formalism
\cite{BV,BV1}. In that formalism, the effective action is BRST invariant by construction,
and thus satisfies
the master equation which provides for the gauge invariance
of the physical sector of the theory \cite{BV,BV1}.

An interest to the gauge dependence problem did appear from the
study of the effective potential,  which appeared to be
gauge-dependent in Yang-Mills theories with spontaneous breaking of
the symmetry, when calculating physically-sensible results (the
energy of the ground state, the masses of the physical particles,
and so on) \cite{Jac, DJac}. In Refs. \cite{Niel,FK} it was
established that the energy of the ground state was
gauge-independent. Later, it was proved \cite{LT3,LT1} that in Yang
-Mills theories the dependence of gauge parameters in the effective
action could be described in terms of gauge-invariant functional
whose arguments (fields) were gauge-dependent (see also recent Refs.
\cite{Niel1,PT} devoted to that problem as resolved via the
procedure of redefinition of the field variables, found in
\cite{LT3,LT1}). Notice that in the general case of gauge theories,
a variation in gauge condition is described in the form of certain
change of the field variables (in terms of anticanonical
transformations) \cite{VLT,BLT-finite}.

Although there are many papers devoted to various aspects of renormalizability of Yang-Mills
theories, gauge dependence of renormalization constants has been studied explicitly
only as for
the gauge field sector \cite{K-SZ}. In the present paper, within the background field
formalism, it is studied
a multiplicative renormalization procedure and gauge dependence as for
Yang-Mills fields interacting
with a multiplet of spinor fields. It is shown that renormalizations
of physical parameters of the
theory are gauge-independent.

The paper is organized as follows. In Section 2, it is discussed the action of
Yang-Mills fields and
spinor fields in the standard approach and in the background field method;
it is also introduced
extended action, which leads in the background field method
to a multiplicative renormalizable
theory of the fields considered; it is also studied the symmetry of the extended action.
In Section 3,
it is established the structure and the arbitrariness is described as for any local
functional with
the quantum numbers of the extended action that satisfies the same set of equations as the
extended action. In Section 4, the equations are derived for the generating functional
of vertexes
(effective action), as a consequence at the quantum level, of the symmetry property of the
extended action; and it is shown that the generating functional of vertexes satisfies
the same
equations as the extended classical action. In Section 5, it is studied the
renormalization procedure
of the theory considered when using the loop expansion technique and the minimal subtraction
scheme; and thus the multiplicative renormalizability of the theory is proved.
In Section 6, the
relations are found between the parameters of the renormalized action and the standard
renormalization constants of fields and vertexes of the interaction,
and renormalized physical
parameters are shown to be gauge-independent. Concluding remarks are given  in Section 7.

Condensed DeWitt's notations \cite{DeWitt} are used through the paper.
Functional derivatives with
respect to field variables are understood as the left. Right derivatives
of a quantity $f$ with respect to the variable $\varphi$
are denoted as $f\frac{\overleftarrow{\delta}}{\delta \varphi}$.

\section{Extended action for Yang-Mills theories}

Let us consider a gauge theory of non-abelian vector fields
$A^{\alpha}_{\mu}=A^{\alpha}_{\mu}(x)$  and spinor fields
$\psi_j=\psi_j\;\!(x), \overline{\psi}_j=\overline{\psi}_j\;\!(x)$ in the $D=4$
Minkowski space-time with the action
\beq
\label{a1}
\mathcal{S}_{YM}(A,\Psi )=\int dx\left( -\frac{1}{4}G_{\mu \nu
}^{\alpha }(A)G_{\mu \nu }^{\alpha }(A)+i\overline{\psi}_j\;\!\gamma
^{\mu }D_{\psi \mu j\;\!k }(A)\psi_k - m\overline{\psi
}_j\;\!\psi_j\right) ,
\eeq
where the notations
\beq
\label{a2}
G_{\mu \nu }^{\alpha }(A)=\pa_{\mu}A^{\alpha}_{\nu}-
\pa_{\nu}A^{\alpha}_{\mu} +gf^{\alpha \beta \gamma }A_{\mu }^{\beta
}A_{\nu }^{\gamma },\quad D_{\psi \mu
j\;\!k}(A)=\pa_{\mu}\delta_{jk}+gt^{\alpha}_{j\;\!k}A^{\alpha}_{\mu},\quad
\Psi=\{\psi,  \overline{\psi}\}
\eeq
are used. In Eqs. (\ref{a1}),
(\ref{a2}) $f^{\alpha\beta\gamma}$ are structure coefficients of a simple compact
gauge Lie group,
$t^{\alpha}=\{t^{\alpha}_{j\;\!k}\}$ are generators of gauge transformations
in sector of spinor fields satisfying the properties,
\beq
\label{a3}
[t^{\alpha},t^{\beta}]=
f^{\alpha\beta\gamma}t^{\gamma},\quad (t^{\gamma})^{+}=-t^{\gamma },
\quad[\gamma^{\mu},t^{\alpha}]=0.
\eeq
Here $\gamma^{\mu}$ are the Dirac matrices,  $g$ and $m$ are
the coupling constant of
gauge interaction and the mass parameter of spinor field, respectively.
 The action
(\ref{a1}) is invariant under gauge transformations with gauge parameters
 $\omega_{\alpha}=\omega_{\alpha}(x)$,
$\delta_{\omega}\mathcal{S}_{YM}(A,\Psi )=0$,
\beq
\nonumber
&&\delta _{\omega }A_{\mu }^{\alpha }= \left( \partial _{\mu }\delta
_{\alpha \beta }+gf^{\alpha \sigma \beta }A_{\mu }^{\sigma }\right)
\omega _{\beta }=
D_{\mu }^{\alpha \beta }(A)\omega_{\beta }, \\
\label{a4}
&&
\delta _{\omega }\psi_j=-gt_{j\;\!k}^{\beta }\psi_{k}\omega _{\beta }, \quad
\delta_{\omega }\overline{\psi}_{j}=
g\overline{\psi}_{k}t_{kj}^{\beta }\;\!\omega _{\beta }.
\eeq
The corresponding Faddeev-Popov action \cite{FP}
$S_{FP}^{(1)}=S_{FP}^{(1)}(A,\Psi, C, \overline{C}, B, \xi)$ in the Feynman gauge
has the form
\beq
\label{a5}
S_{FP}^{(1)}=\mathcal{S}_{YM}(A,\Psi )+\int dx
\Big(\overline{C}^{\alpha}\pa_{\mu}D^{\alpha\beta}_{\mu}(A)C^{\beta}+
B^{\alpha}\pa_{\mu}A^{\alpha}_{\mu}+(\xi/ 2)B^{\alpha}B^{\alpha}\Big),
\eeq
where  $\xi$ is a constant gauge parameter,
$ C^{\alpha}=C^{\alpha}(x), \overline{C}^{\alpha}=\overline{C}^{\alpha}(x)$
represent the Faddeev-Popov ghost fields, $B^{\alpha}=B^{\alpha}(x)$ are auxiliary fields
introducing a gauge fixing condition. The action (\ref{a5}) is invariant under
global supersymmetry (BRST symmetry) \cite{BRS1,T},
\beq
\nonumber
&&\delta_{\lambda} A^{\alpha}_{\mu}=D^{\alpha\beta}_{\mu}(A)C^{\beta}\lambda,\quad
\delta_{\lambda} \psi_j=-gt^{\alpha}_{j\;\!k}\psi_kC^{\alpha}\lambda ,\quad
\delta_{\lambda}  \overline{\psi}_j=g\overline{\psi}_kt^{\alpha}_{kj}C^{\alpha}\lambda,\\
\label{a6}
&&
\delta_{\lambda} C^{\alpha}=\frac{g}{2}
f^{\alpha \beta \gamma }C^{\beta }C^{\gamma }\lambda,\quad
\delta_{\lambda}\overline{C}^{\alpha}=B^{\alpha} \lambda,\quad
\delta_{\lambda} B^{\alpha} =0,
\eeq
where  $\lambda$ is a constant anticommuting parameter.

In the background field formalism \cite{DeW,Abbott} a gauge field
$A^{\alpha}_{\mu}$ entering the classical action  (\ref{a1}) is replaced by
 $A^{\alpha}_{\mu}+{\cal B}^{\alpha}_{\mu}$,
\beq
\label{a7}
\mathcal{S}_{YM}(A,\Psi )\;\rightarrow \;\mathcal{S}_{YM}(A+{\cal B},\Psi ),
\eeq
where
${\cal B}^{\alpha}_{\mu}$ is considered as an external vector field.
 The Faddeev-Popov action is constructed by using
the modified Feynman gauge (the background gauge condition), and reads
\beq
\label{a8}
S_{FP}^{(2)}=\mathcal{S}_{YM}(A+{\cal B},\Psi )+\!\!\int \!dx
\Big(\overline{C}^{\alpha}D^{\alpha\gamma}_{\mu}({\cal B})
D^{\gamma\beta}_{\mu}(A+{\cal B})C^{\beta} +
B^{\alpha}D^{\alpha\beta}_{\mu}({\cal B})A^{\alpha}_{\mu}+(\xi/ 2)B^{\alpha}B^{\alpha}\Big).
\eeq
This action is invariant under BRST transformations of the form
(\ref{a6}) with the following modification of the transformation law
in the gauge field sector,
\beq
\label{a9}
\delta_{\lambda} A^{\alpha}_{\mu}=D^{\alpha\beta}_{\mu}(A+{\cal B})C^{\beta}\lambda.
\eeq

The invariance property of Faddeev-Popov actions (\ref{a5}) and (\ref{a8})
under BRST transformations can be described in the form of non-linear functional equations
for the extended action ${\cal S}_{ext}$ with the help of additional variables
(antifields)
$A^{*\alpha}_{\mu}, \psi^*_j, \overline{\psi}^{*}_j,C^{*\alpha}, \overline{C}^{*\alpha}$,
being sources to the generators of BRST transformations,
\beq
\label{a10}
{\cal S}_{ext}^{(a)}=S_{FP}^{(a)}+\int dx \Big(Q^{*}{\cal R}^{(a)}_{Q}+
\overline{C}^{\;\!*\alpha}B^{\alpha}\Big),\quad a=1,2,
\eeq
where $Q$ means the set of the fields
$\{A^{\alpha}_{\mu}, \psi_j, \overline{\psi}_j,C^{\alpha}\}$ and the symbol
 $Q^*$  is used to indicate the set of the corresponding antifields for fields $Q$,
wherein the BRST transformations (\ref{a4}),
(\ref{a6}) are presented as $\delta_{\lambda}Q={\cal R}^{(a)}_{Q}\lambda$, $a=1,2$.
Then, as a consequence of the BRST symmetry, the actions
${\cal S}_{ext}^{(a)}$ satisfy the master - equation
\beq
\label{a11}
\int dx \Big({\cal S}^{(a)}_{ext}\frac{\overleftarrow{\delta }}{\delta Q}
\frac{\delta }{\delta Q^* }{\cal S}^{(a)}_{ext} +
{\cal S}^{(a)}_{ext}\frac{\overleftarrow{\delta }}{\delta \overline{C} }
\frac{\delta }{\delta \overline{C}^{\;\!*}} {\cal S}^{(a)}_{ext}\Big)=0, \quad a=1,2.
\eeq

To study the structure of renormalizations it is convenient to extend the original set
of the variables with extra fields and auxiliary quantities.
An initial action, we proceed from, when  studying the structure of renormalizations
and dependence of renormalization constants on gauge fixing is the extended action
$S_{ext}=S_{ext}(Q,Q^*,\overline{C}, B, {\cal B},\xi,\theta,\chi)$,
\beq
\nonumber
&&S_{ext}=S_{YM}(A+{\cal B}, \Psi) + \int dx \;Q^{*}{\cal R}_{Q}+\\
\label{a12}
&&
+\int dx\Big(
\overline{C}^{\alpha}D^{\alpha\gamma}_{\mu}({\cal B})
D^{\gamma\beta}_{\mu}(A+{\cal B})C^{\beta} +
B^{\alpha}D^{\alpha\beta}_{\mu}({\cal B})A^{\beta}_{\mu}+
(\xi/ 2)B^{\alpha}B^{\alpha}\Big)+\\
\nonumber && +\int dx \Big(\theta^{\alpha}_{\mu} \lbrack
D^{\alpha\beta}_{\mu}(A+\mathcal{B})\overline{C}^{\beta}- A^{\ast
\alpha}_{\mu}]+\chi [ (A^{\ast \alpha}_{\mu}-
D^{\alpha\beta}_{\mu}(\mathcal{B})\overline{C}^{\beta})A^{\alpha}_{\mu}
+C^{\ast \alpha}C^{\alpha}+\psi ^{\ast }_j\psi_j + \overline{\psi
}^{\ast }_j\overline{\psi }_j]\Big),
\eeq
where $\theta^{\alpha}_{\mu}=\theta^{\alpha}_{\mu}(x)$ are  anticommuting extra
fields and $\chi$ is a constant nilpotent parameter\footnote{These extra variables
have been used first in Ref. \cite{K-SZ}.}.

The action (\ref{a12}) is invariant ($\delta S_{ext}=0$) under the following
transformations of the quantities entered,
\beq
\label{a13} &&\delta A^{\alpha}_{\mu}=
D^{\alpha\beta}_{\mu}(A+\mathcal{B})C^{\beta}+\theta^{\alpha}_{\mu}
=
\frac{\delta }{\delta A^{\ast \alpha}_{\mu}}S_{\mathrm{ext}}+\chi A^{\alpha}_{\mu}, \\
\nonumber
&&\delta B^{\alpha}=-\frac{1}{\xi }\big[D^{\alpha\gamma}_{\mu}(\mathcal{B})
D^{\gamma\beta}(A+\mathcal{B})C^{\beta}+D^{\alpha\beta}_{\mu}(A+
\mathcal{B})\theta^{\beta}_{\mu} +
\chi D^{\alpha\beta}_{\mu}(\mathcal{B})A^{\beta}_{\mu}\big]-\chi B^{\alpha}= \\
\label{a14}
&&\quad\quad =-\frac{1}{\xi }\frac{\delta }{\delta \overline{C}^{\alpha}}
S_{\mathrm{ext}}-\chi B^{\alpha}, \\
\label{a15}
&&\delta C^{\alpha }=\frac{g}{2}f^{\alpha\beta\gamma}C^{\beta}C^{\gamma }=
\frac{\delta }{\delta
C^{\ast \alpha }}S_{\mathrm{ext}}-\chi C^{\alpha }, \\
\label{a16}
&&\delta \overline{C}^{\alpha }=
-\frac{1}{\xi }D^{\alpha\beta}_{\mu}(\mathcal{B})A^{\beta}_{\mu}+
\chi \overline{C}^{\alpha }=-\frac{1}{\xi }\frac{\delta }{\delta B^{\alpha
}}S_{\mathrm{ext}}+B^{\alpha }+\chi \overline{C}^{\alpha }, \\
\label{a17}
&&\delta \psi_j =-gt^{\alpha }_{jk}\psi_k C^{\alpha }=
\frac{\delta }{\delta \psi
^{\ast }_j}S_{\mathrm{ext}}-\chi \psi_j ,\\
\label{a18}
&&
 \delta \overline{\psi }_j=g\overline{
\psi}_kt^{\alpha}_{kj}C^{\alpha}=\frac{\delta}{\delta\overline{\psi}
^{\ast}_j}S_{\mathrm{ext}}-\chi \overline{\psi }_j, \\
\label{a19}
&&\delta A^{\ast \alpha}_{\mu}=\chi A^{\ast \alpha}_{\mu},
\quad \delta C^{\ast \alpha}=-\chi C^{\ast \alpha},\quad
\delta \psi ^{\ast }_j=-\chi \psi ^{\ast }_j,\quad \delta \overline{\psi }^{\ast
}_j=-\chi \overline{\psi }^{\ast }_j, \\
\label{a20}
&&\delta \xi =2\xi \chi ,\quad \delta \mathcal{B}^{\alpha}_{\mu}=-
\theta^{\alpha}_{\mu} ,
\quad \delta \theta^{\alpha}_{\mu}=0,\quad \delta \chi =0.
\eeq
Due to the variations
(\ref{a13}) - (\ref{a20}),
 the invariance condition of the action rewrites
\beq
&&\qquad \int dx\Big( S_{\mathrm{ext}}\frac{\overleftarrow{\delta }}{\delta Q}
\frac{\delta }{\delta Q^{\ast }}S_{\mathrm{ext}}-B\frac{\delta }{\delta
\overline{C}}S_{\mathrm{ext}}-\theta \frac{\delta }{\delta \mathcal{B}}S_{%
\mathrm{ext}}\Big) +  \notag \\
&&+\chi \int dx\Big[Q\frac{\delta}{\delta Q} - Q^*\frac{\delta}{\delta Q^*}
-\overline{C}\frac{\delta }{\delta \overline{C}}-B\frac{%
\delta }{\delta B}
\Big] S_{\mathrm{ext}}+2\chi \xi \frac{\partial }{\partial \xi }S_{\mathrm{%
ext}}=0. \label{a21}
\eeq
Also, the action (\ref{a12}) satisfies the equation
\beq
S_{\mathrm{ext}}\overleftarrow{H^{\alpha }}\omega
_{\alpha }=0, \label{a22}
\eeq
where the notation
\beq
\label{a23}
\overleftarrow{H^{\alpha }}\omega _{\alpha }&=&\int dx\Big\{ \Big[ \frac{%
\overleftarrow{\delta }}{\delta \mathcal{B}_{\mu }^{\beta }}D_{\mu }^{\beta
\alpha }(\mathcal{B})+gf^{\beta \gamma \alpha }\Big( \frac{\overleftarrow{%
\delta }}{\delta A_{\mu }^{\beta }}A_{\mu }^{\gamma }+\frac{\overleftarrow{%
\delta }}{\delta B^{\beta }}B^{\gamma }\Big) + \notag \\
&&+gf^{\beta \gamma \alpha }\Big( \frac{\overleftarrow{\delta }}{\delta
C^{\beta }}C^{\gamma }+\frac{\overleftarrow{\delta }}{\delta \overline{C}%
^{\beta }}\overline{C}^{\gamma }+\frac{\overleftarrow{\delta }}{\delta
A_{\mu }^{\ast \beta }}A_{\mu }^{\ast \gamma }+\frac{\overleftarrow{\delta }%
}{\delta C^{\ast \beta }}C^{\ast \gamma }+\frac{\overleftarrow{\delta }}{%
\delta \theta _{\mu }^{\beta }}\theta _{\mu }^{\gamma }\Big) -  \notag \\
&& -gt_{jk}^{\alpha }\Big( \frac{\overleftarrow{\delta }}{%
\delta \psi _{j}}\psi _{k}+\frac{\overleftarrow{\delta }}{\delta \overline{%
\psi }_{j}^{\ast }}\overline{\psi }_{k}^{\ast }\Big) +g\Big( \frac{%
\overleftarrow{\delta }}{\delta \overline{\psi }_{j}}\overline{\psi }_{k}+%
\frac{\overleftarrow{\delta }}{\delta \psi _{j}^{\ast }}\psi _{k}^{\ast
}\Big) t_{kj}^{\alpha }\Big] \omega _{\alpha }\Big\} ,
\eeq
is used for the operator describing the gauge transformations of the variables
$\mathcal{B}_{\mu }$, $\psi $, $\overline{\psi }$ and simultaneously
the tensor transformation of fields and antifields
$A_{\mu }$, $C$, $\overline{C}$, $B$, $\theta_{\mu} $, $A_{\mu }^{\ast
} $, $\psi ^{\ast }$, $\overline{\psi }^{\ast }$, $C^{\ast }$.
Finally, we notice that the action (\ref{a12}) satisfies the two important relations
linear in  fields $A_{\mu}, B$ and also in derivatives of
variables $B, \overline{C}, A_{\mu }^{\ast}$,
\beq
\label{a24}
&&\frac{\delta }{\delta B^{\alpha }}S_{\mathrm{ext}}=D_{\mu }^{\alpha \beta }(
\mathcal{B})A_{\mu }^{\beta }+\xi B^{\alpha },\\
\label{a25}
&&D_{\mu }^{\alpha \beta }(\mathcal{B})\frac{\delta }{\delta A_{\mu }^{\ast
\beta }}S_{\mathrm{ext}}-\frac{\delta }{\delta \overline{C}^{\alpha }}S_{%
\mathrm{ext}}=-gf^{\alpha \beta \gamma }A_{\mu }^{\beta }\theta _{\mu
}^{\gamma }.
\eeq
The equation (\ref{a25}) means that the action $S_{ext}$ (\ref{a12}) depends
on variables
$A_{\mu}^{\ast \alpha }$ и $\overline{C}^{\alpha }$ in combination $A_{\mu
}^{\ast \alpha }-D_{\mu }^{\alpha \beta }(\mathcal{B})\overline{C}^{\beta }$ only when
$\theta _{\mu }^{\beta }=0$.

We give the table of "quantum"\;\!  numbers of fields, antifields, auxiliary fields and
constant quantities which have been used in constructing $S_{ext}$:
\begin{center}
\begin{tabular}[c]{|c|c|c|c|c|c|c|c|c|c|c|c|c|}
\hline
Quantity & $A,{\cal B}$ & $\psi , \overline{\psi }$ & C,
$\overline{C}$  & B& $\xi$ & $\theta $& $A^*$ &
$\psi^* , \overline{\psi }^*$ & $C^*$ & dx & $\pa_{x}$& $\chi$ \\
\hline
$\varepsilon$ & 0& 1 & 1& 0& 0& 1& 1& 0 & 0& 0 & 0 & 1 \\
\hline
{\rm gh} & 0 & 0 &1, -1 & 0 & 0& 1& -1& -1&-2 &0 &0 & 1\\
\hline
dim &1 & 3/2& 1& 2& 0& 2&2 &3/2 &2 &-4 &1 & 1\\
\hline
$\varepsilon_f$ & 0&1, -1 &0 &0 &0 &0 &0 & -1, 1&0 & 0&0 & 0\\
\hline
\end{tabular}
\end{center}
where $"\varepsilon"$ describes the Grassmann parity,
the symbol $"{\rm gh}"$ is used for the ghost number, $"{\rm dim}"$
denotes the canonical dimension and $"\varepsilon_f"$ means the fermion number.
Using this table of "quantum" \;\!numbers it is easy  to establish quantum numbers
of any quantities found in the text.

\section{General structure of renormalized action}
It is to be proved below that the renormalizable action is a local
functional of field variables, carries the quantum number of the
action ${\cal S}_{ext}$ (\ref{a12}), and satisfies the same
equations (\ref{a21}) - (\ref{a25}) as the action $S_{ext}$. In that
Section we will find the general solution to the equations
(\ref{a21}) - (\ref{a25}) under the extra conditions mentioned.

So, let
\beq
\label{c1}
P=\int dx P(x),
\eeq
where $P(x)$ is a local
polynomial in all variables $Q,Q^*,\overline{C}, B, {\cal
B},\xi,\theta,\chi$ with ${\rm dim}(P(x))=4$. Require the functional
$P$ to satisfy the equations (\ref{a21}) - (\ref{a25}) with
substitution  $S_{ext} \to P$, and  let  $P$  be of the form
\beq
\label{c2}
P=P_{00}+P^{(1)}+\chi P^{(2)}, \eeq where \beq \label{c3}
&&P_{00}=\int dx\left( B^{\alpha}D^{\alpha\beta}_{\mu}(\mathcal{B})
A^{\beta}_{\mu}+\frac{\xi }{2}B^{\alpha}B^{\alpha}+
g\theta_{\mu}^{\alpha}f^{\alpha\beta\gamma}A_{\mu}^{\beta}
\overline{C}^{\gamma}\right) ,  \\
\label{c3a} &&\varepsilon (P^{(1)})=0,\quad {\rm
gh}(P^{(1)})=0,\quad {\rm dim}(P^{(1)})=0,\quad
\varepsilon_f((P^{(1)}))=0,\\
\label{c4} &&\varepsilon (P^{(2)})=1,\quad {\rm
gh}(P^{(2)})=-1,\quad {\rm dim}(P^{(2)})=-1,\quad
\varepsilon_f((P^{(2)}))=0,
\eeq
and the functionals  $P^{(1)}$ and
$P^{(2)}$ do not depend on $\chi$. It follows from the equation
(\ref{a24}) for $P$, and representation  (\ref{c3}) that $P^{(1)}$
and $P^{(2)}$ do not depend on $B^{\alpha}$, не зависят,
\beq
\label{c5}
 P^{(k)}=P^{(k)}(Q,Q^{\ast },\overline{C},\mathcal{B},\xi ,\theta ),\ k=1,2,
\eeq
By introducing new variables $\mathcal{A}_{\mu }^{\ast \alpha
}(x)$,
\beq
\label{c6}
\mathcal{A}_{\mu }^{\ast \alpha }=A_{\mu
}^{\ast \alpha }-D_{\mu }^{\alpha \beta
}(\mathcal{B})\overline{C}^{\beta },\ \mathcal{A}_{\mu }^{\alpha
}=A_{\mu }^{\alpha },
\eeq
we define new functionals
$\tilde{P}^{(k)}$ by the rule
\beq
\label{c7}
\tilde{P}^{(k)}=\tilde{P}^{(k)}(\Omega,\Omega^{*},\mathcal{B},\overline{C},
\xi,\theta)=P^{(k)}(Q,Q^*,\mathcal{B},\overline{C},\xi,\theta)|_{A^*\to
\mathcal{A}^*+D(\mathcal{B})\overline{C}},\quad k=1,2,
\eeq
to find that $\tilde{P}^{(k)}$ do not depend on the fields
$\overline{C}^{\;\alpha}$,
\beq
\label{c8}
\tilde{P}^{(k)}=\tilde{P}^{(k)}(\Omega,\Omega^{\ast},\mathcal{B},\xi,\theta).
\eeq
In the relations (\ref{c7}) and (\ref{c8}) the following notations
\beq
\label{c9}
\Omega =\{\mathcal{A},\psi,\overline{\psi},C\},\ \mathcal{A}=A,\
\Omega^{\ast}=\{\mathcal{A}^{\ast},\psi^{\;\ast},\overline{\psi}^{\ast},
C^{\ast}\}
\eeq
are used.

Independence of functionals $\tilde{P}^{(k)}$ of the fields  $\overline{C}^{\;\alpha}$
and relations
\beq
\nonumber
&&P^{(k)}\int dx\Big[ \frac{\overleftarrow{\delta }}{\delta \mathcal{B}%
_{\mu }^{\beta }}D_{\mu }^{\beta \alpha }(\mathcal{B})+gf^{\beta \gamma
\alpha }\Big( \frac{\overleftarrow{\delta }}{\delta \overline{C}^{\beta }}%
\overline{C}^{\gamma }+\frac{\overleftarrow{\delta }}{\delta A_{\mu }^{\ast
\beta }}A_{\mu }^{\ast \gamma }\Big) \Big] = \\
\label{c10}
&&\ =\tilde{P}^{(k)}\int dx\Big[ \frac{\overleftarrow{\delta }}{\delta
\mathcal{B}_{\mu }^{\beta }}D_{\mu }^{\beta \alpha }(\mathcal{B})+gf^{\beta
\gamma \alpha }\frac{\overleftarrow{\delta }}{\delta \mathcal{A}_{\mu
}^{\ast \beta }}\mathcal{A}_{\mu }^{\ast \gamma }\Big] ,
\eeq
allow one to write down the following set of equations as for $\tilde{P}^{(k)}$,
\beq
\label{c11}
&&\qquad \qquad \int dx\Big[ \tilde{P}^{(1)}\frac{\overleftarrow{\delta }}{\delta
\Omega}\frac{\delta }{\delta \Omega^{\ast }}\tilde{P}%
^{(1)}-\theta _{\mu }^{\alpha }
\frac{\delta }{\delta \mathcal{B}_{\mu}^{\alpha }}\tilde{P}^{(1)}\Big] =0, \\
\nonumber
&&2\xi \frac{\partial }{\partial \xi }\tilde{P}^{(1)}=\int dx\Big[
\tilde{P}^{(1)}\Big( \frac{\overleftarrow{\delta }}{\delta \Omega}
\frac{\delta }{\delta \Omega^{\ast }}-\frac{\overleftarrow{\delta }}{%
\delta \Omega^{\ast }}\frac{\delta }{\delta \Omega}\Big)
\tilde{P}^{(2)}-\theta _{\mu }^{\alpha }\frac{\delta }{\delta \mathcal{B%
}_{\mu }^{\alpha }}\tilde{P}^{(2)}\Big] + \\
&&\qquad \qquad\quad +\int dx\Big[ \Big(\Omega^*\frac{\delta}{\delta \Omega^*}
-\Omega\frac{\delta}{\delta \Omega}\Big)\tilde{P}^{(1)}\Big ],
\label{c12}
\eeq
\beq
\label{c13}
\tilde{P}^{(k)}\overleftarrow{\tilde{h}^{\alpha}}\omega_{\alpha}=0,\ k=1,2,
\eeq
where
\beq
&&\overleftarrow{\tilde{h}^{\alpha }}\omega _{\alpha }=\int dx\Big\{ %
\Big[ \frac{\overleftarrow{\delta }}{\delta \mathcal{B}_{\mu }^{\beta }}%
D_{\mu }^{\beta \alpha }(\mathcal{B})+gf^{\beta \gamma \alpha }\Big( \frac{%
\overleftarrow{\delta }}{\delta \mathcal{A}_{\mu }^{\beta }}\mathcal{A}%
_{m|\mu }^{\gamma }+\frac{\overleftarrow{\delta }}{\delta C^{\beta }}%
C^{\gamma }\Big) +  \notag \\
&&\qquad\qquad+gf^{\beta \gamma \alpha }\Big( \frac{\overleftarrow{\delta }}{\delta
\mathcal{A}_{\mu }^{\ast \beta }}\mathcal{A}_{\mu }^{\ast \gamma }+\frac{%
\overleftarrow{\delta }}{\delta C^{\ast \beta }}C^{\ast \gamma }+\frac{%
\overleftarrow{\delta }}{\delta \theta^{\beta }}\theta _{\mu
}^{\gamma }\Big) -  \notag \\
&&\qquad -gt_{jk}^{\alpha }\Big( \frac{\overleftarrow{\delta }}{%
\delta \psi _{j}}\psi _{k}+\frac{\overleftarrow{\delta }}{\delta
\overline{\psi }_{j}^{\ast }}\overline{\psi }_{k}^{\ast }\Big) +g\Big(
\frac{\overleftarrow{\delta }}{\delta \overline{\psi }_{j}}\overline{\psi }%
_{k}+\frac{\overleftarrow{\delta }}{\delta \psi _{j}^{\ast }}\psi
_{k}^{\ast }\Big) t_{kj}^{\alpha }\Big] \omega _{\alpha }\Big\}.
\label{c14}
\eeq

When studying the structure of functionals and further investigating
it appears  useful a consequence of the equation
(\ref{c13})  at $\omega_{\alpha}={\rm const}$,
\beq
\label{c15}
\tilde{P}^{(k)}\overleftarrow{T^{\alpha }}=0,\ k=1,2,
\eeq
where
\beq
&&\overleftarrow{T^{\alpha }}=\int dx\Big\{ f^{\beta \gamma \alpha
}\Big( \frac{\overleftarrow{\delta }}{\delta \mathcal{B}_{\mu }^{\beta }}%
\mathcal{B}_{\mu }^{\gamma }+\frac{\overleftarrow{\delta }}{\delta \mathcal{A%
}_{\mu }^{\beta }}\mathcal{A}_{\mu }^{\gamma }+\frac{\overleftarrow{%
\delta }}{\delta C^{\beta }}C^{\gamma }\Big) + \notag \\
&&\qquad\qquad+f^{\beta \gamma \alpha }\Big( \frac{\overleftarrow{\delta }}{\delta
\mathcal{A}_{\mu }^{\ast \beta }}\mathcal{A}_{\mu }^{\ast \gamma }+\frac{%
\overleftarrow{\delta }}{\delta C^{\ast \beta }}C^{\ast \gamma }+\frac{%
\overleftarrow{\delta }}{\delta \theta _{\mu }^{\beta }}\theta _{\mu
}^{\gamma }\Big) -  \notag \\
&&\qquad -t_{jk}^{\alpha }\Big( \frac{\overleftarrow{\delta }}{\delta \psi
_{j}}\psi _{k}+\frac{\overleftarrow{\delta }}{\delta \overline{\psi }%
_{j}^{\ast }}\overline{\psi }_{k}^{\ast }\Big) +\Big( \frac{%
\overleftarrow{\delta }}{\delta \overline{\psi }_{j}}\overline{\psi }%
_{k}+\frac{\overleftarrow{\delta }}{\delta \psi _{j}^{\ast }}\psi
_{k}^{\ast }\Big) t_{kj}^{\alpha }\Big\} .
\label{c16}
\eeq
We refer to equations of the form (\ref{c15}) as the ones  of
the $T$-symmetry for the corresponding functional.

Using the properties of the functional $\tilde{P}^{(2)}$ (\ref{c4}), its locality as well as
axial symmetry, Poincare - and $T$- symmetries we find the general
representation,
\beq
\label{c17}
\tilde{P}^{(2)}=
\int dx\left[ Z_{1}\mathcal{A}^{\ast \alpha}_{\mu}\mathcal{A}^{\alpha}_{\mu}
+Z_{2}C^{\ast \alpha}C^{\alpha}+Z_{3}\psi ^{\ast }_j\psi_j +
Z_{4}\overline{\psi }^{\;\ast }_j
\overline{\psi }_j+Z_{1}^{\prime}\mathcal{A}^{\ast \alpha}_{\mu}
\mathcal{B}^{\alpha}_{\mu}\right],
\eeq
where $Z_i, i=1,2,3,4,$ и $Z_{1}^{\prime }$ are arbitrary constants.
Further, when using the equation (\ref{c13}) for $\tilde{P}^{(2)}$ we get that
  $Z_{1}^{\prime }=0$. The final expression for  $\tilde{P}^{(2)}$ has the form
\beq
\label{c18}
\tilde{P}^{(2)}=
\int dx\left[ Z_{1}\mathcal{A}^{\ast \alpha}_{\mu}\mathcal{A}^{\alpha}_{\mu}
+Z_{2}C^{\ast \alpha}C^{\alpha}+Z_{3}\psi ^{\ast }_j\psi_j+
Z_{4}\overline{\psi }^{\;\ast }_j
\overline{\psi }_j \right].
\eeq
Notice that the functional $\tilde{P}^{(2)}$ does not depend on the fields
$\theta^{\alpha}_{\mu} $. By taking ({\ref{c18}}) into account
 the equation  (\ref{c12}) reduces to the following one
\beq
\nonumber
&&2\xi \frac{\partial }{\partial \xi }\tilde{P}^{(1)}=
\int dx\Big[( Z_1-1)\Big(\mathcal{A}^{\alpha}_{\mu}
\frac{\delta}{\delta\mathcal{A}^{\alpha}_{\mu}}-
\mathcal{A}^{*\alpha}_{\mu}\frac{\delta}{\delta\mathcal{A}^{*\alpha}_{\mu}}\Big)+
(Z_2-1)\Big(C^{\alpha}\frac{\delta}{\delta C^{\alpha}}-
C^{*\alpha}\frac{\delta}{\delta C^{*\alpha}}\Big)+\\
&&
\qquad\qquad\quad +(Z_3-1)\Big(\psi_j\frac{\delta}{\delta\psi_j}-
{\psi }^{\;\ast }_j\frac{\delta}{\delta{\psi }^{\;\ast }_j}\Big)+
(Z_4-1)\Big(\overline{\psi }_j\frac{\delta}{\delta\overline{\psi }_j}-
\overline{\psi}^{\;\ast}_j\frac{\delta}{\delta\overline{\psi}^{\;\ast}_j}\Big)
\Big]\tilde{P}^{(1)},
\label{c19}
\eeq
describing the dependence of renormalization constants on the gauge parameter $\xi$.
We refer to the equation ({\ref{c11}}) as an extended master-equation and to ({\ref{c19}})
as a gauge dependence equation.

\subsection{Solution to the extended master-equation}

Now we consider a solution to the extended master-equation ({\ref{c11}})
for the functional $\tilde{P}^{(1)}$ as presented it in the form
\beq
\label{d1}
\tilde{P}^{(1)}=\tilde{P}_{\theta }^{(1)}+ \tilde{P}_{\Omega^*}^{(1)}+\tilde{P}_{\psi
}^{(1)}+ \tilde{P}_{\mathcal{AB}}^{(1)}.
\eeq
The functional $\tilde{P}_{\theta }^{(1)}$ rewrites  as
\beq
\label{d2}
\tilde{P}_{\theta }^{(1)}=\int dx\;
\theta^{\alpha}_{\mu}(x)\tilde{P}^{\alpha}_{\mu\theta }(x) ,
\eeq
and the functionals $\tilde{P}_{\mathcal{AB}}^{(1)}$, $\tilde{P}_{\psi
}^{(1)}$, $\tilde{P}_{\Omega^*}^{(1)}$ do not depend on the fields
$\theta^{\alpha}_{\mu}$. By taking into account the properties  ${\rm
dim}(\tilde{P}^{\alpha}_{\mu\theta })=2$, ${\rm
gh}(\tilde{P}^{\alpha}_{\mu\theta })=-1$, $\varepsilon
(\tilde{P}^{\alpha}_{\mu\theta })=1$,
$\varepsilon_{f}(\tilde{P}^{\alpha}_{\mu\theta })=0$, as well as the Poincare - and  $T$
- symmetries of the functional $\tilde{P}_{\theta }^{(1)}$, we find that
\beq
\label{d3}
\tilde{P}^{\alpha}_{\mu\theta
}=-Z_{5}\mathcal{A}^{\ast\alpha }_{\mu},\quad
\tilde{P}_{\theta}^{(1)}=-Z_{5}\int dx \; \theta^{\alpha}_{\mu}(x)
\mathcal{A}^{\ast \alpha}_{\mu}(x)=-Z_{5}\int dx \;
\theta^{\alpha}_{\mu} \mathcal{A}^{\ast \alpha}_{\mu},
\eeq
where $Z_5$ is an arbitrary constant.

The functional $\tilde{P}_{\Omega^*}^{(1)}$ is linear in the antifields  $\Omega^*$
(\ref{c9}), and the functionals  $\tilde{P}_{\mathcal{AB}}^{(1)}$ and
$\tilde{P}_{\psi }^{(1)}$ do not depend on the antifields $\Omega^*$. The functional
$\tilde{P}_{\Omega^*}^{(1)}$ can be represented in the form
\beq
\label{d4}
\tilde{P}_{\Omega ^{\ast }}^{(1)}=\tilde{P}_{\mathcal{A}%
^{\ast }}^{(1)}+\tilde{P}_{C^{\ast }}^{(1)}+\tilde{P}_{\psi ^{\ast }}^{(1)}+%
\tilde{P}_{\overline{\psi }^{\ast }}^{(1)}.
\eeq
By using the  arguments analogous  to those  led us to  the structure of the functional
$\tilde{P}_{\theta }^{(1)}$ (\ref{d3}), we obtain
\beq
\label{d5}
&&\tilde{P}_{\mathcal{A}^{\ast }}^{(1)}=\int dx\left[ Z_{6}\mathcal{A}^
{\ast\alpha}_{\mu}D^{\alpha\beta}_{\mu}(\mathcal{B)}C^{\beta}+
gZ_{7\beta \gamma }^{\alpha }\mathcal{A}^{\ast \alpha }_{\mu}
\mathcal{A}^{\beta }_{\mu}C^{\gamma }+gZ_{7\beta \gamma }^{'\alpha }
\mathcal{A}^{\ast \alpha }_{\mu}
\mathcal{B}^{\beta }_{\mu}C^{\gamma }\right] ,  \\
\label{d6}
&&\tilde{P}_{C^{\ast }}^{(1)}=\int
dx\; \frac{g}{2}Z_{8\beta \gamma }^{\alpha }
C^{\ast \alpha }C^{\beta }C^{\gamma },  \\
\label{d7}
&&\tilde{P}_{\psi ^{\ast }}^{(1)}=-\int dx\;
gZ_{9jk}^{\alpha } \psi
_{j}^{\ast }\psi _{k}C^{\alpha },  \\
\label{d8}
&&\tilde{P}_{\overline{\psi }^{\ast }}^{(1)}=\int dx\;
gZ_{10kj}^{\alpha }
 \overline{\psi }_{j}^{\ast }\overline{\psi }_{k}C^{\alpha }.
\eeq
Taking into account the gauge symmetry in the external field ${\cal B}$ (see the equation
(\ref{c13})), we find that $Z_{7\beta \gamma }^{'\alpha }=0$.
The quantities $"Z"$ introduced in (\ref{d5}) - (\ref{d8}) are constants
that satisfy the equations,
\beq
\label{d9}
&&F_{\gamma\delta }^{\alpha }Z_{7\delta \sigma }^{\beta }-Z_{7\gamma \delta }
^{\beta}F_{\delta\sigma}^{\alpha}=f^{\alpha\beta\lambda}
Z_{7\gamma\sigma}^{\lambda},\\
\label{d10}
&&F_{\gamma \delta
}^{\alpha }Z_{8\delta \sigma }^{\beta }-Z_{8\gamma \delta }^{\beta
}F_{\delta \sigma }^{\alpha }=f^{\alpha \beta \delta }Z_{8\gamma \sigma }^{\delta },\\
\label{d11}
&&t_{jl}^{\alpha }Z_{9lk}^{\beta }-Z_{9jl}^{\beta }t_{lk}^{\alpha
}=f^{\alpha \beta \gamma }Z_{9jk}^{\gamma },\\
\label{d12}
&&t_{kl}^{\alpha }Z_{10lj}^{\beta
}-Z_{10kl}^{\beta }t_{lj}^{\alpha }=f^{\alpha \beta \gamma }Z_{10kj}^{\gamma }.
\eeq
Notice that if  $Z_{7\gamma \sigma }^{\lambda }=Z_{7}f^{\gamma \lambda \sigma }$,
$Z_{8\gamma \sigma }^{\lambda }=Z_{8}f^{\gamma \lambda \sigma }$,
$Z_{9jk}^{\alpha }=Z_{9}t_{jk}^{\alpha }$, $Z_{10jk}^{\alpha }=Z_{10}t_{jk}^{\alpha }$,
then the corresponding equations (\ref{d9}) - (\ref{d12}) hold and the functionals
$\tilde{P}_{\mathcal{A}^{\ast }}^{(1)}, \tilde{P}_{C^{\ast }}^{(1)},
\tilde{P}_{\psi ^{\ast }}^{(1)}, \tilde{P}_{\overline{\psi }^{\ast }}^{(1)}$
(\ref{d5}) - ( \ref{d8})  satisfy the equation (\ref{c13})  by themselves .

In its turn, taking into account the axial symmetry,
the Poincare - and the  $T$- invariance we determine the general structure of the functional
$\tilde{P}_{\psi }^{(1)}$,
\beq
\nonumber
&&\tilde{P}_{\psi }^{(1)}=\int dx\big[ iZ_{11}\overline{\psi
}_{j}\gamma
^{\mu }D_{\psi \mu }(\mathcal{B})\psi _{j}+igZ_{11jk}^{\prime \alpha }%
\overline{\psi }_{j}\gamma ^{\mu }\mathcal{B}_{\mu }^{\alpha }\psi_{k}+\\
\label{d13}
&&\qquad\qquad +igZ_{12jk}^{\alpha }\overline{\psi }_{j}\gamma ^{\mu }\mathcal{A}_{\mu
}^{\alpha }\psi _{k}-mZ_{13}\overline{\psi }_{j}\psi _{j}\big],
\eeq
where constants  $Z_{12jk}^{\alpha }$ satisfy the equations
\beq
\label{d14}
t_{jl}^{\alpha }Z_{12lk}^{\beta }-Z_{12jl}^{\beta }
t_{lk}^{\alpha}=f^{\alpha \beta \gamma }Z_{12jk}^{\gamma }.
\eeq
The contribution to the
$\tilde{P}_{\psi }^{(1)}\overleftarrow{\tilde{h}^{\alpha }}\omega
_{\alpha }$, proportional to  $\partial _{\mu }\omega ^{\alpha }$, has the form
\beq
\label{d15}
ig\overline{\psi }_{j}Z_{11jk}^{\prime \alpha }\gamma ^{\mu }\psi
_{k}\partial _{\mu }\omega ^{\alpha },
\eeq
so that it follows from the equation  (\ref{c13}) that the equalities
$Z_{11jk}^{\prime \alpha }=0$ and
\beq
\label{d16}
\tilde{P}_{\psi }^{(1)}=\int dx\left[ iZ_{11}\overline{\psi }\gamma ^{\mu
}D_{\psi \mu }(\mathcal{B})\psi +igZ_{12jk}^{\alpha }\overline{\psi }%
_{j}\gamma ^{\mu }\mathcal{A}_{\mu }^{\alpha }\psi _{k}-mZ_{13}\overline{%
\psi }_{j}\psi _{j}\right]
\eeq
hold. Notice that in the case $Z_{12jk}^{\alpha }=Z_{12}t_{kj}^{\alpha }$ the equations
(\ref{d14}) are fulfilled and the functional $\tilde{P}_{\psi }^{(1)}$ (\ref{d16})
satisfies the equation (\ref{c13}).

Insert  the representation for the functional $\tilde{P}^{(1)}$ in
the form (\ref{d1}) into the equation (\ref{c11}). Then, analysis of
the $\theta \psi \overline{\psi }$ components in the extended
master-equation (\ref{c11}) yields \beq \label{d17} Z_{12jk}^{\alpha
}=Z_{12}t_{jk}^{\alpha },\quad Z_{12}=Z_{11}/Z_{5}, \eeq and   the
possibility to represent the functional  $\tilde{P}_{\psi }^{(1)}$
as \beq \label{d18} \tilde{P}_{\psi }^{(1)}=\int dx\left[
iZ_{11}\overline{\psi }_{j}\gamma ^{\mu }D_{\psi \mu jk}(U)\psi
_{k}-mZ_{13}\overline{\psi }_{j}\psi _{j}\right], \eeq where the
notation
\beq
\label{d19}
U=\{U^{\alpha}_{\mu}\}, \quad
U^{\alpha}_{\mu}=
Z_{5}^{-1}\mathcal{A}^{\alpha}_{\mu}+\mathcal{B}^{\alpha}_{\mu}
\eeq
is used. The $\theta \mathcal{A}^{\ast }C$ components  in the
equation (\ref{c11})lead to the relations
\beq
\label{d20}
Z_{7\beta
\gamma}^{\alpha}=Z_7f^{\alpha\beta\gamma},\quad Z_{7}=
\frac{Z_{6}}{Z_{5}},
\eeq
and to the representation
\beq
\label{d21}
\tilde{P}_{\mathcal{A}^{\ast }}^{(1)}=\int dx\; Z_{6}
\mathcal{A}^{\ast \alpha}_{\mu}D^{\alpha\beta}_{\mu}(U)C^{\beta}.
\eeq
Consideration of the $\mathcal{A}^{\ast }\mathcal{A}CC$
components in the equation (\ref{c11}) gives the relations
\beq
\label{d22} Z_{8\beta \gamma }^{\alpha }=Z_{8}f^{\alpha \beta \gamma
},\quad Z_{8}=Z_{7}=\frac{Z_{6}}{Z_{5}},
\eeq
and the representation
for the functional  $\tilde{P}_{C^{\ast }}^{(1)}$ in the form
\beq
\label{d23}
\tilde{P}_{C^{\ast }}^{(1)}=\int dx\;
\frac{g}{2}\frac{Z_{6}}{Z_{5}}f^{\alpha \beta \gamma } C^{\ast
\alpha }C^{\beta }C^{\gamma }.
\eeq
Studying the
$\overline{\psi}\psi\partial C$, $\overline{\psi}\psi \mathcal{B}C$
and  $m\overline{\psi }\psi C$ components in the equations
(\ref{c11}) lead to the relations
\beq
\label{d24}
Z_{10jk}^{\alpha
}=Z_{9jk}^{\alpha },\quad Z_{9jk}^{\alpha }= Z_{9}t_{jk}^{\alpha
},\quad Z_{9}=\frac{Z_{6}}{Z_{5}},\quad Z_{10jk}^{\alpha
}=\frac{Z_{6}}{Z_{5}}t_{jk}^{\alpha },
\eeq
and, as a consequence,
to the representation of the functionals
 $\tilde{P}_{\psi ^{\ast }}^{(1)}$ and
$\tilde{P}_{\overline{\psi }^{\;\ast }}^{(1)}$  as
\beq
\label{d25}
&&\tilde{P}_{\psi ^{\ast }}^{(1)}=-\int dx\; g\frac{Z_{6}}{Z_{5}}\; \psi
_{j}^{\ast }t_{jk}^{\alpha }\psi _{k}C^{\alpha },\\
\label{d26}
&&\tilde{P}_{\overline{\psi }^{\;\ast }}^{(1)}=\int
dx\; g\frac{Z_{6}}{Z_{5}}\; \overline{\psi }_{j}^{\;\ast}t_{kj}^{\alpha}
\overline{\psi}_{k}C^{\alpha} =\int dx\; g\frac{Z_{6}}{Z_{5}}\;
\overline{\psi}_{j}t_{jk}^{\alpha}\overline{\psi}_{k}^{\;\ast}C^{\alpha}.
\eeq

The functional $\tilde{P}_{\mathcal{AB}}^{(1)}$ depends on the fields ${\cal A}$
and ${\cal B}$ only. The $\theta \mathcal{AB}$ components   in the equation (\ref{c11})
allow us to conclude that the functional  $\tilde{P}_{\mathcal{AB}}^{(1)}$
 depends on the fields
${\cal A}$ and  ${\cal B}$ only in combination (\ref{d19}),
\beq
\label{d27}
\tilde{P}_{\mathcal{AB}}^{(1)}(\mathcal{A},\mathcal{B})=X(U).
\eeq
Finally, consideration of the $\mathcal{AB}C$ components in the equation (\ref{c11})
leads to equations for the functional  $X(U)$ (\ref{d27})
\beq
\label{d28}
D_{\mu }^{\alpha \beta }(U)\frac{\delta }{\delta U_{\mu }^{\beta }(x)}
X(U)=0.
\eeq
The required solution to the equations (\ref{d28}) can be written in the form
\beq
\label{d29}
\tilde{P}_{\mathcal{AB}}^{(1)}(\mathcal{A},\mathcal{B})= X(U)=-\int dx\;\frac{
1}{4}Z_{14}\;G_{\mu \nu }^{\alpha }( U) G_{\mu \nu
}^{\alpha }( U ),
\eeq
where
\beq
\label{d30}
G_{\mu \nu }^{\alpha }( U) =\partial _{\mu }U_{\nu }^{\alpha
}-\partial _{\nu }U_{\mu }^{\alpha }+gf^{\alpha \beta \gamma }U_{\mu
}^{\beta }U_{\nu }^{\gamma }.
\eeq
Thus the general solution to the extended master-equation, $\tilde{P}^{(1)}$, is constructed.
It is defined by fifth independent arbitrary constants  $Z_5, Z_6, Z_{11}, Z_{13}, Z_{14}$
and has the form
\beq
\nonumber
&&\tilde{P}^{(1)}=\int dx \Big[-\frac{
1}{4}Z_{14}\;G_{\mu \nu }^{\alpha }( U) G_{\mu \nu
}^{\alpha }(U) + iZ_{11}\overline{\psi }_{j}\gamma
^{\mu }D_{\psi \mu jk}(U)\psi _{k}-mZ_{13}\overline{\psi }_{j}\psi
_{j}-\\
\nonumber
&&\qquad\qquad\qquad -Z_{5} \theta^{\alpha}_{\mu}\mathcal{A}^{\ast \alpha}_{\mu}+
Z_{6}
\mathcal{A}^{\ast \alpha}_{\mu}D^{\alpha\beta}_{\mu}(U)C^{\beta}+\\
&&
\qquad\qquad\qquad +g\frac{Z_{6}}{Z_{5}}\left(f^{\alpha \beta
\gamma } C^{\ast \alpha }C^{\beta }C^{\gamma }+
\overline{\psi }_{j}^{\;\ast }t_{kj}^{\alpha }\overline{\psi }%
_{k}C^{\alpha }-\psi _{j}^{\ast }t_{jk}^{\alpha }\psi _{k}C^{\alpha
}\right) \Big]. \label{d31}
\eeq

Notice that at $Z_1=Z_2= Z_3=Z_4=Z_5=Z_6= Z_{11}=Z_{13}= Z_{14}=1$,
the equality (the initial condition),
\beq \label{d32}
P_{Z=1}=S_{ext},
\eeq
holds where
$S_{ext}$  is given by the formula (\ref{a12}).

\subsection{Solution to the gauge dependence equation}

Consider now a solution to the equation (\ref{c19}) describing the gauge dependence
of the constants entering the general solution constructed, $\tilde{P}^{(1)}$, to the
extended master-equation (\ref{d31}). By studying
the $\mathcal{A}^{\ast }\theta $, $\mathcal{A}D(U)G(U)$, $\mathcal{A}%
^{\ast }f\mathcal{A}C$ and $\overline{\psi }\gamma t\mathcal{A}\psi $ structures
 in the equation (\ref{c19}), we derive the following relation
\beq
\label{e1b}
2\xi\dot{Z}_{5}=-(Z_{1}-1)Z_{5}\ \Rightarrow\ Z_{1}=1-2\xi\frac{\dot{Z}_{5}}{Z_5}.
\eeq
Henceforth  we use the notation
\beq
\label{e2b}
\dot{I}\equiv \frac{\partial }{\partial \xi }I ,
\eeq
for any quantity $I=I(\xi ,...)$ depending on the gauge parameter $\xi$.

Analysis of the $\mathcal{A}^{\ast }D(U)C$ components  in the equation
(\ref{c19}) gives the relation
\beq
\label{e3b}
2\xi\dot{Z}_6=(Z_2-Z_1)Z_6\ \Rightarrow\ Z_2=1
+2\xi\Big(\frac{\dot{Z}_6}{Z_6}-\frac{\dot{Z}_5}{Z_5}\Big).
\eeq
Considering the $\overline{\psi}\gamma D_{\psi}(U)\psi$ components in the equation
(\ref{c19}), we obtain
\beq
\label{e4a}
2\xi \dot{Z}_{11}=(Z_{3}+Z_{4}-2)Z_{11}.
\eeq
Analyzing the $m\overline{\psi}\psi$ components  in the equation  (\ref{c19}), we find
\beq
\label{e5aa}
2\xi \dot{Z}_{13}=(Z_{3}+Z_{4}-2)Z_{13}.
\eeq
By making use of  the change of constants $"Z"$
\beq
\label{e5ab}
Z_{13}=Z_{11}Z_{15},\quad Z_3-Z_4=2Z_{16},
\eeq
the equations  (\ref{e4a}), (\ref{e5aa}) rewrite in the form
\beq
\label{e5a}
\dot{Z}_{15}=0,\quad Z_3=1+\xi\frac{\dot{Z}_{11}}{Z_{11}}+Z_{16},\quad
Z_4=1+\xi\frac{\dot{Z}_{11}}{Z_{11}}-Z_{16}.
\eeq
Finally, consideration of the $G(U)G(U)$ components
 in the equation  (\ref{c19}) leads to the important statement that,
\beq
\label{e6}
\dot{Z}_{14}=0.
\eeq
Analysis of the $\psi ^{\ast }t\psi C$, $\overline{\psi }^{\ast }t^{t}
\overline{\psi }C$ and $C^{\ast }fCC$ components
 in the equation (\ref{c19}) gives no new information.

Below, in Section 5 we find that all constants $"Z"$ can be interpreted
as renormalization constants which are uniquely defined
from the conditions of divergence elimination.

Let us formulate the results obtained in that Section in the form of a lemma.

{\it Lemma}: Let
\beq
\label{e7}
P=\int dx P(Q,Q^*,\overline{C}, B,
{\cal B},\xi,\theta,\chi),
\eeq
be a local functional of all variables, obey the quantum numbers of the action $S_{ext}$
and satisfy all equations (\ref{a21}) - (\ref{a25}) as well as extra symmetries
(Poincare-invariance and so on) which have been used in solving the equations
(\ref{a21}) - (\ref{a25}) with substitution $S_{ext} \to P$.

Then the functional  $P$ has the form
\beq
\label{e8a}
P=P_{00}+P^{(1)}+\chi P^{(2)},
\eeq
where $P_{00}$ is given by the formula
(\ref{c3}), $P^{(1)}$ and  $P^{(2)}$ do not depend on $B^\alpha$ and
$\chi$ and are functionals of arguments  $\Omega$, $\Omega^{*}$,
$\mathcal{B}$, $\overline{C}$, $\xi$, $\theta$,
\beq
\label{e9}
 P^{(k)}=P^{(k)}(Q,Q^{\ast},\overline{C},\mathcal{B},\xi,\theta)=
\tilde{P}^{(k)}=\tilde{P}^{(k)}(\Omega,\Omega^{*},\mathcal{B},\overline{C},
\xi,\theta),\ k=1,2,
\eeq
\beq
\Omega =\{\mathcal{A},\psi,\overline{\psi},C\},\
\Omega^{\ast}=\{\mathcal{A}^{\ast},\psi^{\;\ast},\overline{\psi}^{\ast},
C^{\ast}\},
\eeq
\beq
\mathcal{A}_{\mu }^{\ast \alpha }=A_{\mu }^{\ast \alpha }-D_{\mu }^{\alpha
\beta }(\mathcal{B})\overline{C}^{\beta },\ \mathcal{A}_{\mu }^{\alpha
}=A_{\mu }^{\alpha }.
\eeq
The functionals  $\tilde{P}^{(k)}$ read
\beq
\nonumber
&&\tilde{P}^{(1)}=\int dx\Big[-\frac{1}{4}Z_{14}\;G_{\mu \nu }^{\alpha}(U)
G_{\mu\nu}^{\alpha}(U) + iZ_{11}\overline{\psi}_{j}\gamma^{\mu}
D_{\psi\mu jk}(U)\psi_k-mZ_{13}\overline{\psi}_j\psi_j-
Z_5\theta^{\alpha}_{\mu}\mathcal{A}^{\ast\alpha}_{\mu}+\\
&&\qquad\qquad
+Z_{6}\mathcal{A}^{\ast\alpha}_{\mu}D^{\alpha\beta}_{\mu}(U)C^{\beta}+
g\frac{Z_{6}}{Z_{5}}\left(f^{\alpha\beta\gamma}C^{\ast\alpha}C^{\beta}C^{\gamma}+
\overline{\psi}_{j}^{\;\ast}t_{kj}^{\alpha}\overline{\psi}_{k}C^{\alpha
}- \psi_{j}^{\ast}t_{jk}^{\alpha }\psi_{k}C^{\alpha}\right) \Big],
\label{e10a} \eeq \beq \label{e11a} \tilde{P}^{(2)}= \int dx\left[
Z_{1}\mathcal{A}^{\ast \alpha}_{\mu}\mathcal{A}^{\alpha}_{\mu}
+Z_{2}C^{\ast \alpha}C^{\alpha}+Z_{3}\psi ^{\ast }_j\psi_j+
Z_{4}\overline{\psi }^{\;\ast }_j \overline{\psi }_j \right]. \eeq
\beq \label{e12a} U=\{U^{\alpha}_{\mu}\}, \quad U^{\alpha}_{\mu}=
Z_{5}^{-1}\mathcal{A}^{\alpha}_{\mu}+\mathcal{B}^{\alpha}_{\mu},
\eeq
\beq
\label{e13}
Z_{1}\!=\!1-2\xi\frac{\dot{Z}_{5}}{Z_5},\quad
\!\!Z_2\!=\!1+2\xi\Big
(\frac{\dot{Z}_6}{Z_6}-\frac{\dot{Z}_5}{Z_5}\Big),\quad
\!\!Z_3\!=\!1+\xi \frac{\dot{Z}_{11}}{Z_{11}}+Z_{16},\quad \!\!
Z_4\!=\!1+\xi\frac{\dot{Z}_{11}}{Z_{11}}-Z_{16},
\eeq
\beq
\label{e14} Z_{13}=Z_{11}Z_{15},\quad \dot{Z}_{14}=0,\quad
\dot{Z}_{15}=0,
\eeq
where  $Z_5,Z_6,Z_{11},Z_{16}$ are arbitrary constants  depending perhaps on  $\xi$, and
$Z_{14},Z_{15}$ are arbitrary constants not depending on $\xi$.

The inverse statement, being perhaps  trivial but nevertheless important,  is true:
if the functional $P$ has the form  (\ref{e8a}), (\ref{e9}), (\ref{e10a}),
(\ref{e11a}) and the relations  (\ref{e12a}), (\ref{e13}),
(\ref{e14}) are fulfilled then this functional satisfies the equations
(\ref{a21}) - (\ref{a25}).

\section{Generating functional of vertex functions}
\noindent
It is convenient to define the generating functional of Green functions by making
use of the action functional $P$ constructed in the previous Section as the action yields
then  a finite theory certainly. In what follows we re-denote the functional
$P$, $P\equiv S_R$, and, respectively, $P^{(k)}\equiv S_R^{(k)}$,
$\tilde{P}^{(k)}\equiv\tilde{S}_R^{(k)}, k=1,2$.

The generating functional of Green functions is given
by the  functional integral,
\beq
\label{b1}
Z(J_{\Phi},L)=\int
d\Phi\exp\Big(\frac{i}{\eta}\big[S_R+J_{\Phi}\Phi\big]\Big)
=\exp\Big\{\frac{i}{\eta }W(J_{\Phi},L)\Big\},
\eeq
with $\eta$ standing for a parameter of a loop expansion as to the  expression in the exponential
in (\ref{b1}),  $W(J_{\Phi},L)$ that is
the generating functional of connected Green functions, and the notations are introduced
$\Phi=\{Q,\overline{C},B\}$ и
$L=\{L^A\}=\{\mathcal{B},Q^{\ast},\xi,\theta,\chi\}$, and $J_{\Phi}$ as
for the sources to the fields $\Phi$. Also, we assume that all the constants
$"Z"$ are functions of  $\eta$, $"Z"$=$"Z"(\eta)$, expandable  in Taylor power series,
$Z_i(0)=1,\; \dot{Z}_i=O(\eta),\; i=5,6,11,14,15$, $Z_{16}=O(\eta)$.
In that case the functional $S_R$ becomes a function of $\eta$,
\beq
\label{b1a}
S_R=S_R(\eta)=\sum_{l=0}^{\infty}\eta^lS_{R,l},\quad
S_R^{[k]}=\sum_{l=0}^k\eta^lS_{R,l},
\eeq
so that  all the functionals
$S_{R,l}$ are linear combination of a single set of monomials,
\beq
\label{b1b}
S_{R,l} =\sum_{i=1}^Ia_{l,i}S_i,
\eeq
where  $\{S_i,i=1,...,I\}$ is a sub-set of monomials
which the action $S_{ext}$ is expanded in, and  $a_{l,i}$ are constant coefficients
for $l$-loop order.

The generating functional of vertex Green functions (effective action)
is defined by the Legendre transformation
\beq
\label{b2}
\Gamma(\Phi_{m|},L)=W(J_{\Phi},L)-J_{\Phi}\Phi_{m|},\quad  \Phi_{m|}=
\frac{\delta}{\delta J_{\Phi}}W(J_{\Phi},L),\
\eeq
has the quantum numbers $\varepsilon(\Gamma)=0$, ${\rm gh}(\Gamma)=0$,
${\rm dim}(\Gamma)=0$,  $\varepsilon_f(\Gamma)=0$, and satisfies the relations
\beq
\label{b3}
\Gamma (\Phi_{m|},L)\frac{\overleftarrow{\delta }}{\delta
\Phi_{m|}}=-J_{\Phi}(\Phi_{m|},L),\quad \Gamma
(\Phi_{m|},L)\frac{\overleftarrow{\delta }}{\delta L^{A}
}=W(J_{\Phi},L)\frac{\overleftarrow{\delta }}{\delta
L^{A}}\;.
\eeq

Functional average of the equations (\ref{a21}) - (\ref{a25}) with substitution
$S_{ext} \to S_R$ yields  the corresponding equations for the functional
$\Gamma=\Gamma(\Phi_{m|},L)$, copying the equations for $S_R$,
\beq
&&\int dx\Big( \Gamma \frac{\overleftarrow{\delta }}{\delta Q_{m|}}\frac{
\delta }{\delta Q^{\ast }}\Gamma -B_{m|}\frac{\delta }{\delta \overline{C}
_{m|}}\Gamma -\theta \frac{\delta }{\delta \mathcal{B}}\Gamma \Big) +2\chi
\xi \frac{\partial }{\partial \xi }\Gamma +  \notag \\
\label{b4}
&&+\chi \int dx\Big[ \Big(Q_{m|}\frac{\delta}{\delta Q_{m|}} -
Q^*\frac{\delta}{\delta Q^*}-\overline{
C}_{m|}\frac{\delta }{\delta \overline{C}_{m|}}-B_{m|}\frac{\delta }{\delta
B_{m|}}\Big) \Gamma \Big] =0,
\eeq
\beq
\label{b5}
\Gamma \overleftarrow{H_{m|}^{\alpha }}\omega _{\alpha }=0,
\eeq
where $\overleftarrow{H_{m|}^{\alpha }}\omega _{\alpha }$ is given by the expression
(\ref{a23}) with the replacement $\Phi$ $\rightarrow \Phi_{m|}$,
\beq
\label{b6}
&&\frac{\delta }{\delta B_{m|}^{\alpha }}\Gamma =D_{\mu }^{\alpha \beta }(%
\mathcal{B})A_{m|\mu }^{\beta }+\xi B_{m|}^{\alpha },  \\
\label{b7}
&&D_{\mu }^{\alpha \beta }(\mathcal{B})\frac{\delta }{\delta A_{\mu }^{\ast
\beta }}\Gamma -\frac{\delta }{\delta \overline{C}_{m|}^{\alpha }}\Gamma
=-gf^{\alpha \beta \gamma }A_{m|\mu }^{\beta }\theta _{\mu }^{\gamma }
\eeq

Represent the functional $\Gamma$ in the following form
\beq
\label{b8}
\Gamma=\Gamma _{00}+\Gamma^{(1)}+\chi \Gamma^{(2)},
\eeq
where
\beq
\label{b9}
\Gamma _{00}=\int dx\Big(B_{m|}D(\mathcal{B})A_{m|}+\frac{\xi}{2}%
B_{m|}^{2}+g\theta_{\mu}^{\alpha}f^{\alpha\beta\gamma}A_{m|\mu}^{\beta}
\overline{C}_{m|}^{\gamma}\Big),
\eeq
and the functionals   $\Gamma ^{(1)}$ and $\Gamma ^{(2)}$ do not depend on the parameter
$\chi $.
Due to the structure chosen for the functional
(\ref{b9}) it follows from the equations  (\ref{b6})
and (\ref{b7}) that the functionals
 $\Gamma ^{(1)}$ and $\Gamma ^{(2)}$ do not depend on the fields  $B^{\alpha}_{m|}$,
\beq
\label{b10}
\frac{\delta }{\delta B_{m|}}\Gamma ^{(k)}=0,\ \Gamma ^{(k)}=
\Gamma^{(k)}(Q_{m|},\overline{C}_{m|},\mathcal{B},Q^{\ast},\xi ,\theta),\ k=1,2,
\eeq
and satisfy the equations
\beq
\label{b11}
\Big(D_{\mu}^{\alpha\beta}(\mathcal{B})\frac{\delta}{\delta A_{\mu}^{\ast\beta}}
-\frac{\delta }{\delta \overline{C}_{m|}^{\alpha }}\Big)
\Gamma ^{(k)}=0,\ k=1,2,
\eeq
In its turn, the equation (\ref{b4}) splits in the two,
one of which  is closed as for the functional $\Gamma^{(1)}$,
\beq
\label{b12}
\int dx\Big[ \Gamma ^{(1)}\frac{\overleftarrow{\delta }}{\delta Q_{m|}}%
\frac{\delta }{\delta Q^{\ast }}\Gamma ^{(1)}-g\theta _{\mu }^{\alpha
}f^{\alpha \beta \gamma }\overline{C}_{m|}^{\beta }\frac{\delta }{\delta
A_{\mu }^{\ast \gamma }}\Gamma ^{(1)}-\theta _{\mu }^{\alpha }\frac{\delta }{%
\delta \mathcal{B}_{\mu }^{\alpha }}\Gamma ^{(1)}\Big] =0,
\eeq
and the second  includes both the functionals and describes their dependence on the gauge
parameter $\xi$,
\beq
&&2\xi\frac{\partial}{\partial\xi}\Gamma^{(1)}=\int dx\Big[\Gamma^{(1)}
\Big(\frac{\overleftarrow{\delta}}{\delta Q_{m|}}\frac{\delta}
{\delta Q^{\ast}}-\frac{\overleftarrow{\delta}}{\delta Q^{\ast}}\frac{\delta}
{\delta Q_{m|}}\Big)\Gamma^{(2)}-\Big(g\theta_{\mu}^{\alpha}f^{\alpha\beta
\gamma}\overline{C}_{m|}^{\beta}\frac{\delta}{\delta A_{\mu }^{\ast\gamma}}+
\theta_{\mu}^{\alpha}\frac{\delta}{\delta\mathcal{B}_{\mu}^{\alpha}}\Big)
\Gamma^{(2)}\Big]+ \notag \\
\label{b13}
&&\qquad\qquad +\int dx\Big[\Big(\overline{C}_{m|}\frac{\delta}{\delta
\overline{C}_{m|}}-Q_{m|}\frac{\delta}{\delta Q_{m|}}+Q^*\frac{\delta}
{\delta Q^*}\Big) \Gamma ^{(1)}\Big] .
\eeq
The equation (\ref{b5}) rewrites now in the form of the two equations
as for the functionals  $\Gamma ^{(1)}$ and $\Gamma ^{(2)}$,
\beq
\label{b14}
\Gamma ^{(k)}\overleftarrow{h_{m|}^{\alpha }}\omega _{\alpha }=0,\ k=1,2,
\eeq
where
\beq
\label{b15}
&&\overleftarrow{h_{m|}^{\alpha }}\omega _{\alpha }=\int dx\Big\{ \Big[
\frac{\overleftarrow{\delta }}{\delta \mathcal{B}_{\mu }^{\beta }}D_{\mu
}^{\beta \alpha }(\mathcal{B})+gf^{\beta \gamma \alpha }\Big( \frac{%
\overleftarrow{\delta }}{\delta A_{m|\mu }^{\beta }}A_{m|\mu }^{\gamma }+%
\frac{\overleftarrow{\delta }}{\delta C_{m|}^{\beta }}C_{m|}^{\gamma
}\Big)  +  \notag \\
&&\qquad +gf^{\beta \gamma \alpha }\Big( \frac{\overleftarrow{\delta }}{\delta
\overline{C}_{m|}^{\beta }}\overline{C}_{m|}^{\gamma }+\frac{\overleftarrow{%
\delta }}{\delta A_{\mu }^{\ast \beta }}A_{\mu }^{\ast \gamma }+\frac{%
\overleftarrow{\delta }}{\delta C^{\ast \beta }}C^{\ast \gamma }+\frac{%
\overleftarrow{\delta }}{\delta \theta _{\mu }^{\beta }}\theta _{\mu
}^{\gamma }\Big) -  \notag \\
&&\qquad -gt_{jk}^{\alpha }\Big( \frac{\overleftarrow{\delta }}{%
\delta \psi _{m|j}}\psi _{m|k}+\frac{\overleftarrow{\delta }}{\delta
\overline{\psi }_{j}^{\ast }}\overline{\psi }_{k}^{\ast }\Big) +g\Big(
\frac{\overleftarrow{\delta }}{\delta \overline{\psi }_{m|j}}\overline{\psi }%
_{m|k}+\frac{\overleftarrow{\delta }}{\delta \psi _{j}^{\ast }}\psi
_{k}^{\ast }\Big) t_{kj}^{\alpha }\Big] \omega _{\alpha }\Big\} .
\eeq

As for the equations (\ref{b11}), it is convenient to introduce the variables
$\mathcal{A}_{\mu }^{\ast \alpha }=\mathcal{A}_{\mu }^{\ast \alpha }(x)$,
$\mathcal{A}_{m|\mu }^{\ast \alpha }=\mathcal{A}_{m|\mu }^{\ast \alpha }(x)$
\beq
\label{b16}
\mathcal{A}_{\mu}^{\ast\alpha}=A_{\mu}^{\ast\alpha}-D_{\mu}^{\alpha\beta}
(\mathcal{B})\overline{C}^{\beta},\quad\mathcal{A}_{m|\mu}^{\ast\alpha}=
A_{\mu}^{\ast\alpha}-D_{\mu}^{\alpha\beta}(\mathcal{B})
\overline{C}_{m|}^{\beta},
\eeq
and to use the following convention
\beq
\label{b17}
\mathcal{A}_{\mu }^{\alpha }=A_{\mu }^{\alpha },
\eeq
as for the sake of uniformity. Also, introduce  the new functionals
 $\tilde{\Gamma}^{(k)}$ by the rule,
\beq
\label{b18}
\tilde{\Gamma}^{(k)}(\mathcal{B},\overline{C}_{m|},\mathcal{A}_{m|}^{\ast},
\Lambda_{m|})=\Gamma^{(k)}(\mathcal{B},\overline{C}_{m|},A^{\ast},
\Lambda_{m|})\big|_{A^{\ast}\to\mathcal{A}_{m|}^{\ast}+D_{\mu}^{\alpha\beta}
(\mathcal{B})\overline{C}_{m|}^{\beta}},
\eeq
where the notation
\beq
\label{b19}
\Lambda=\{Q,\psi^{\ast},\overline{\psi}^{\ast},C^{\ast},\xi,\theta\},\quad
\Lambda_{m|}=\{Q_{m|},\psi^{\ast},\overline{\psi}^{\ast},C^{\ast},\xi,\theta\}
\eeq
is used. With the definitions (\ref{b16}) - (\ref{b18}) taken into account, we have
\beq
\label{b20}
&&\frac{\delta }{\delta A_{\mu }^{\ast \alpha }}\Gamma ^{(k)}=\frac{\delta }{%
\delta \mathcal{A}_{m|\mu }^{\ast \alpha }}\tilde{\Gamma}^{(k)},
 \\
\label{b21}
&&\frac{\delta }{\delta \overline{C}_{m|}^{\alpha }}\Gamma ^{(k)}=\frac{%
\delta }{\delta \overline{C}_{m|}^{\alpha }}\tilde{\Gamma}^{(k)}+D_{\mu
}^{\alpha \beta }(\mathcal{B})\frac{\delta }{\delta \mathcal{A}_{m|\mu
}^{\ast \beta }}\tilde{\Gamma}^{(k)},   \\
\label{b22}
&&\frac{\delta }{\delta \mathcal{B}_{\mu }^{\alpha }}\Gamma ^{(k)}=\frac{%
\delta }{\delta \mathcal{B}_{\mu }^{\alpha }}\tilde{\Gamma}^{(k)}-gf^{\alpha
\beta \gamma }\overline{C}_{m|}^{\beta }\frac{\delta }{\delta \mathcal{A}%
_{m|\mu }^{\ast \gamma }}\tilde{\Gamma}^{(k)},\quad k=1,2.
\eeq
Then, we find from the equations (\ref{b11}), (\ref{b18}),  (\ref{b20}), (\ref{b21})
that
\beq
\label{b23}
\frac{\delta }{\delta \overline{C}_{m|}^{\alpha }}\tilde{\Gamma}^{(k)}=0,
\eeq
the functionals  $\tilde{\Gamma}^{(k)}, k=1,2$, do not depend on the fields
$\overline{C}_{m|}^{\alpha}$,
\beq
\label{b24}
\tilde{\Gamma}^{(k)}=\tilde{\Gamma}^{(k)}(\Omega
_{m|},\Omega _{m|}^{\ast },\mathcal{B}, \xi ,\theta) .
\eeq
Henceforth we  use the notations
\beq
\label{b25}
\Omega_{m|}=\{\mathcal{A}_{m|},\psi_{m|},\overline{\psi}_{m|},C_{m|}\},\quad
\mathcal{A}=A,\quad\Omega_{m|}^{\ast}=\{\mathcal{A}_{m|}^{\ast},\psi^{\ast},
\overline{\psi}^{\ast},C^{\ast}\},\quad k=1,2.
\eeq
Now, with  (\ref{b18}), (\ref{b24}), (\ref{b25}) taken into account, the ones (\ref{b12}),
(\ref{b13}) rewrite in the form
\beq
\label{b26}
&&\qquad\qquad
\frac{1}{2}(\tilde{\Gamma}^{(1)},\tilde{\Gamma}^{(1)})-\int dx\Big(\theta\frac{\delta}
{\delta\mathcal{B}}\Big)\tilde{\Gamma}^{(1)}=0, \\
&&2\xi\frac{\partial}{\partial\xi}\tilde{\Gamma}^{(1)}=(\tilde{\Gamma}^{(1)}
,\tilde{\Gamma}^{(2)})+\int dx\Big(\Omega_{m|}^{\ast}\frac{\delta}
{\delta\Omega_{m|}^{\ast}}-\Omega_{m|}\frac{\delta}{\delta\Omega_{m|}}\Big)
\tilde{\Gamma}^{(1)}-\int dx\Big(\theta\frac{\delta}{\delta \mathcal{B}}\Big)
\tilde{\Gamma}^{(2)},
\label{b27}
\eeq
where the notation for the antibracket  \cite{BV,BV1} is used,
\beq
\label{b27a}
&&(F,G)=\frac{1}{2}F\int dx\Big(\frac{\overleftarrow{\delta}}{\delta\Omega_{m|}}
\frac{\delta}{\delta\Omega_{m|}^{\ast}}-\frac{\overleftarrow{\delta}}
{\delta\Omega_{m|}^{\ast}}\frac{\delta}{\delta\Omega_{m|}}\Big)G.
\eeq
Further, with the relations (\ref{b18}), (\ref{b24}) and
\beq
\nonumber
&&\Gamma ^{(k)}\int dx\Big[ \frac{\overleftarrow{\delta }}{\delta \mathcal{B%
}_{\mu }^{\beta }}D_{\mu }^{\beta \alpha }(\mathcal{B})+gf^{\beta \gamma
\alpha }\Big( \frac{\overleftarrow{\delta }}{\delta \overline{C}%
_{m|}^{\beta }}\overline{C}_{m|}^{\gamma }+\frac{\overleftarrow{\delta }}{%
\delta A_{\mu }^{\ast \beta }}A_{\mu }^{\ast \gamma }\Big) \Big] = \\
\label{b29}
&&\qquad =\tilde{\Gamma}^{(k)}\int dx\Big[ \frac{\overleftarrow{\delta }}{\delta
\mathcal{B}_{\mu }^{\beta }}D_{\mu }^{\beta \alpha }(\mathcal{B})+gf^{\beta
\gamma \alpha }\frac{\overleftarrow{\delta }}{\delta \mathcal{A}_{m|\mu
}^{\ast \beta }}\mathcal{A}_{m|\mu }^{\ast \gamma }\Big]
\eeq
we find that
\beq
\label{b30}
\Gamma^{(k)}\overleftarrow{h_{m|}^{\alpha}}\omega_{\alpha}=\tilde{\Gamma}^{(k)}
\overleftarrow{\tilde{h}_{m|}^{\alpha}}\omega_{\alpha}=0,\ k=1,2,
\eeq
where the operator $\overleftarrow{\tilde{h}_{m|}^{\alpha}}\omega_{\alpha}$ is defined in the
equality (\ref{c14}) with the replacement
$\Omega \rightarrow \Omega_{m|}, \;\Omega^*\rightarrow \Omega^*_{m|}$.

Then,  when studying the tensor structure of divergence parts of
the generating functional of vertexes, it is convenient to use a consequence
of the equations (\ref{b30}) in particular case $\omega _{\alpha }(x)=\mathrm{const}$, i.e.
as to a global $T_{m|}$-symmetry:
\beq
\label{b32}
\tilde{\Gamma}^{(k)}\overleftarrow{T_{m|}^{\alpha }}=0,\ k=1,2,
\eeq
where the operators  $\overleftarrow{T_{m|}^{\alpha }}$ are defined by the equations
(\ref{c16}) with the replacement
$\Omega \rightarrow \Omega_{m|}, \; \Omega^*\rightarrow \Omega^*_{m|}$.

\section{Renormalization}
\noindent
In that Section we study the structure of renormalizations, and show
the multiplicative character of the renormalizabiliuty of the model
considered. The main role  in that study is given to resolving  the extended
master-equation (\ref{c11}) and the one (\ref{c19}) describing the gauge dependence.
We show that the renormalized quantum action and the effective action
satisfy exactly their master equations to each subsequent order in  loops.
In  this resolving, the structure of the renormalized quantum action is
determined by the same monomials in fields and antifields as it does for
the non-renormalized quantum action with constants determined by the
divergencies of the effective action.
For the sake of notational simplicity, we omit lower case $m|$ of any arguments
of any functionals.

\subsection{Tree approximation ($\eta =0$)}
Consider the tree approximation for the functional $\Gamma$,
$\Gamma_{0}=S_{\mathrm{ext}}$, written in new variables as
\begin{eqnarray}
&&\Gamma_{0}=\Gamma _{00}+\Gamma_{0}^{(1)}+\chi\Gamma_{0}^{(2)}, \\
&&\Gamma _{0}^{(1)}=\tilde{\Gamma}_{0}^{(1)},\quad
\Gamma_{0}^{(2)}=\tilde{\Gamma}_{0}^{(2)},
\end{eqnarray}
where
\beq
\label{c3aa}
&&\Gamma_{00}=\int dx\left( B^{\alpha}D^{\alpha\beta}_{\mu}(\mathcal{B})
A^{\beta}_{\mu}+\frac{\xi }{2}B^{\alpha}B^{\alpha}+
g\theta_{\mu}^{\alpha}f^{\alpha\beta\gamma}A_{\mu}^{\beta}
\overline{C}^{\gamma}\right).
\eeq
Represent the functional $\tilde\Gamma^{(1)}_0$ in the form
\beq
\nonumber
&&\qquad\qquad\tilde\Gamma^{(1)}_0=\Gamma_{0\theta}+
\Gamma_{0\Omega^*}+\Gamma_{0\psi} +\Gamma_{0\mathcal{AB}},  \\
\label{da18}
&&\Gamma_{0\psi}=\Gamma_{0\psi|1}+\Gamma_{0\psi|2},\quad
\Gamma_{0\Omega^*}=
\Gamma_{0\mathcal{A}^{\ast}}+\Gamma_{0C^{\ast}}+
\Gamma_{0\psi^{\ast}}+\Gamma_{0\overline{\psi}^{\ast}},
\eeq
where the following notations being further useful
\beq
\label{da19}
&&\Gamma_{0\theta}=\int
dx\mathcal{A}^{\ast\alpha}_{\mu}\theta^{\alpha}_{\mu},\\
\label{da20}
&&\Gamma_{0\mathcal{A}^{\ast}}=\int dx\mathcal{A}^{\ast\alpha}_{\mu}
D^{\alpha\beta}_{\mu}(\mathcal{A}+\mathcal{B})C^{\beta}, \\
\label{da21}
&&\Gamma_{0C^{\ast}}=\int dx\frac{g}{2}f^{\alpha\beta\gamma}
C^{\ast\alpha}C^{\beta}C^{\gamma},\\
\label{da22}
&&\Gamma_{0\psi^{\ast}}=-\int dx g\psi_j^{\ast}t_{j\;\!k}^{\alpha}
\psi_{k}C^{\alpha}, \;
\Gamma_{0\overline{\psi}^{\ast}}=\int dx g\overline{\psi}_{j}^{\ast}
t^{\alpha}_{kj}\overline{\psi}_{k}C^{\alpha}, \\
\label{da24}
&&\Gamma_{0\psi|1}=\int dx\left[i\overline{\psi}\gamma^{\mu}D_{\psi\mu}
(\mathcal{A}+\mathcal{B})\psi\right], \quad
\Gamma_{0\psi|2}=-m\int dx\overline{\psi}\;\!\psi, \\
\label{da25}
&&\Gamma_{0\mathcal{AB}}=-\int dx\;\frac{
1}{4}\;G_{\mu \nu }^{\alpha }( \mathcal{A}+\mathcal{B}) G_{\mu \nu
}^{\alpha }( \mathcal{A}+\mathcal{B} ),
\eeq
are introduced. In its turn, the functional $\tilde{\Gamma}_0^{(2)}$ has the form
\beq
\label{c18aa}
\tilde{\Gamma}_0^{(2)}=
\int dx\left[\mathcal{A}^{\ast\alpha}_{\mu}\mathcal{A}^{\alpha}_{\mu}
+C^{\ast\alpha}C^{\alpha}+\psi^{\ast}_j\psi_j+
\overline{\psi}^{\;\ast }_j\overline{\psi}_j\right].
\eeq
Remind that the functional $\Gamma_0$ satisfies the equations
(\ref{a21}) - (\ref{a25}).

\subsection{(l+1)-loop approximation}

We carry out the proof of the multiplicative renormalizability via the mathematical induction
method in the framework of loop expansion of the effective action with the use of
the minimal subtraction scheme. To this end we suppose that we managed
to find such parameters
$Z_i^{[l]}$,
\beq
\nonumber
&&Z_i^{[l]}=\sum_{n=0}^l\eta^n z_{i,n},\quad i=5,6,11,14,15,16,\quad
\dot{Z}_{14}^{[l]}=\dot{Z}_{15}^{[l]}=0, \\
\nonumber
&&1-2\xi\frac{\dot{Z}_5^{[l]}}{Z_5^{[l]}}=Z_1^{[l]}+O(\eta^{l+1}),\quad
1+2\xi\Big(\frac{\dot{Z}_6^{[l]}}{Z_6^{[l]}}-\frac{\dot{Z}_5^{[l]}}
{Z_5^{[l]}}\Big)=Z_2^{[l]}+O(\eta^{l+1}), \\
\label{ca1}
&&1+\xi\frac{\dot{Z}_{11}^{[l]}}
{Z_{11}^{[l]}}+Z_{16}^{[l]}=Z_3^{[l]}+O(\eta^{l+1}),\quad
1+\xi\frac{\dot{Z}_{11}^{[l]}}{Z_{11}^{[l]}}-Z_{16}^{[l]}=Z_4^{[l]}+
O(\eta^{l+1}),
\eeq
that the  $l$-loop approximation for $\Gamma$,
$\Gamma^{[l]}=\sum_{n=0}^l \eta^n\Gamma_n$, is a finite functional.
We are to show that it is possible to pick up the $l+1$-loop approximation for $Z_i$,
\beq
Z_i=Z_i^{[l]}+z_{i,l+1}+O(\eta^{l+2}), \quad i=5,6,11,14,15,16,
\quad {\dot z}_{14,l+1}={\dot z}_{15,l+1}=0,
\eeq
which does compensate the  divergences of
$l+1$-loop approximation for the functional $\Gamma$.

Represent the action $S_R$ in the form
\beq
S_R=S_R^{[l]}+\eta^{l+1}s_{l+1}+O(\eta^{l+2}),
\eeq
where  $S_R^{[l]}$
is the action $S_R$ with independent parameters  $Z_i$ replaced by
$Z_i^{[l]}$, and satisfying the equations (\ref{a21}) - (\ref{a25}),
and the functional  $s_{l+1}$ reads
\beq
s_{l+1}=s_{l+1}^{(1)}+\chi s_{l+1}^{(2)}.
\eeq
For the functional $s_{l+1}^{(1)}$ we use the representation
\beq
\label{ca4}
s_{l+1}^{(1)}=s_{\theta,l+1}+s_{\Omega^*,l+1}+s_{\psi,l+1}+
s_{\mathcal{AB},l+1},
\eeq
where
\beq
\label{ca4a}
&&s_{\theta,l+1}=z_{5,l+1}\Gamma_{0\theta}, \\
\label{ca4b}
&&s_{\mathcal{A}^*,l+1}=z_{6,l+1}\Gamma_{0\mathcal{A}^*}-
z_{5,l+1}\mathcal{A}\pa_{\mathcal{A}}\Gamma_{0\mathcal{A}^*}, \\
&&s_{C^*,l+1}=(z_{6,l+1}-z_{5,l+1})\Gamma_{0C^*}, \\
&&s_{\psi^*,l+1}=(z_{6,l+1}-z_{5,l+1})\Gamma_{0\psi^*}, \;
s_{\overline{\psi}^{\;\ast},l+1}=(z_{6,l+1}-z_{5,l+1})
\Gamma_{0\overline{\psi}^{\;\ast}}, \\
&&s_{\psi,l+1}=z_{11,l+1}\Gamma_{0\psi|1}-z_{5,l+1}\mathcal{A}
\pa_{\mathcal{A}}\Gamma_{0\psi|1}+(z_{11,l+1}+z_{15,l+1})\Gamma_{0\psi|2}, \\
&&s_{\mathcal{AB},l+1}=z_{14,l+1}\Gamma_{0\mathcal{AB}}-
z_{5,l+1}\mathcal{A}\pa_{\mathcal{A}}\Gamma_{0\mathcal{AB}}.
\eeq
In its turn the functional  $s_{l+1}^{(2)}$ has the form
\beq
\nonumber
&&s_{l+1}^{(2)}=\int dx\Big[2\xi\dot{z}_{5,l+1}\mathcal{A}^{\ast}\mathcal{A}+
2\xi(\dot{z}_{6,l+1}-\dot{z}_{5,l+1})C^{\ast}C+  \\
\label{ca3}
&&\qquad\qquad+\xi(\dot{z}_{11,l+1}+
z_{16,l+1})\psi^{\ast}\psi+ +\xi(\dot{z}_{11,l+1}-z_{16,l+1})\overline{\psi}^{\;\ast}
\overline{\psi}\Big].
\eeq
Here (and below in this Section) we use the abbreviation to denote
the variational derivatives  of the kind
\beq
\frac{\delta}{\delta\mathcal{A}}\to\pa_{\mathcal{A}},\quad
\mathcal{A}\pa_{\mathcal{A}}=\int dx\; \mathcal{A}\frac{\delta}{\delta\mathcal{A}},
\eeq
when it does not cause an ambiguity.

Let us  study the structure of the functional $\Gamma$ with the accuracy including
the $(l+1)$-loop approximation. It is described by the diagrams
with vertexes from the action $S_R$ with parameters
$z_{i,n}$, $i=5,6,11,14,15,16$, $0\leq n\leq l+1$, or, in other words,
by vertexes from the action $S_R^{[l]}$ and from the action
$s_{l+1}$. As we are interested in diagrams of the loop order not higher than $l+1$,
the vertexes from $s_{l+1}$  cannot appear in loop diagrams, i.e. vertexes from
$s_{l+1}$ give the "tree" contribution to $\Gamma$, equal to $\eta^{l+1}s_{m|,l+1}$.
Other diagrams are generated by the action $S_R^{[l]}$. Let $\Gamma(S_R^{[l]})$
be the contribution of thouse diagrams into the functional $\Gamma$, i.e.
\beq
\label{ca5}
\Gamma=\Gamma(S_R^{[l]})+\eta^{l+1}s_{m|,l+1}+O(\eta^{l+2}).
\eeq

As the functional $S_R^{[l]}$ satisfies the required equations, the functional
 $\Gamma(S_R^{[l]})$ satisfies  the same equations with the replacement
  $Q,\overline{C},B\rightarrow Q_{m|},\overline{C}_{m|},B_{m|}$.

Represent the functional  $\Gamma(S_R^{[l]})$ in the form
\beq
\label{ca6}
\Gamma(S_R^{[l]})=\Gamma_{00}+\Gamma^{(1)}(S_R^{[l]})+
\chi\Gamma^{(2)}(S_R^{[l]}),
\eeq
By repeating the calculations of Section 3 we find that
\beq
\label{ca7}
\Gamma^{(k)}(S_R^{[l]}|Q_{m|},\overline{C}_{m|},\mathcal{B},Q^{\ast},B_{m|},
\xi,\theta)=\tilde{\Gamma}^{(k)}(S_R^{[l]}|\Omega_{m|},\Omega_{m|}^{\ast},
\mathcal{B},\xi,\theta),\ k=1,2,
\eeq
and the functionals  $\tilde{\Gamma}^{(k)}(S_R^{[l]})$ satisfy the equations
(\ref{b26}), (\ref{b27}) и (\ref{b30}).

Represent the functionals  $\tilde{\Gamma}^{(k)}(S_R^{[l]})$ in the
form of sums of divergent and finite (after removing a
regularization) parts. Taking into account that the functionals
$\tilde{\Gamma}^{(k)}(S_R^{[l]})$ are, by assumption, finite to the
$n$-loop approximations, $0\leq n\leq l$, we obtain
\beq
\label{ca8}
\tilde{\Gamma}^{(k)}(S_R^{[l]})=\tilde{\Gamma}^{(k)}(S_R^{[l]})_{\mathrm{fin}}+
\eta^{l+1}\tilde{\Gamma}^{(k)}(S_R^{[l]})_{l+1,\mathrm{div}}+O(\eta^{l+2}),
\eeq
\beq
\label{ca9}
&&\Gamma=\Gamma(S_R^{[l]})_{\mathrm{fin}}+\eta^{l+1}\left[\Gamma(S_R^{[l]})_
{l+1,\mathrm{div}}+s_{l+1}\right]+O(\eta^{l+2}), \\
\label{ca9a}
&&\Gamma(S_R^{[l]})_{l+1,\mathrm{div}}=\tilde{\Gamma}^{(1)}(S_R^{[l]})_{l+1,
\mathrm{div}}+\chi\tilde{\Gamma}^{(2)}(S_R^{[l]})_{l+1,\mathrm{div}},
\eeq
so that the functionals $\tilde{\Gamma}^{(k)}(S_R^{[l]})_{l+1,\mathrm{div}}$
are local ones of arguments with the quantum numbers of the action $S_{ext}$
and contain divergent terms only (the minimal subtraction scheme). Then, as a consequence of
the equations (\ref{b26}), (\ref{b27}) и (\ref{b30}), they satisfy the following equations,
\beq
\label{ca10}
&&(\tilde{\Gamma}_0,\tilde{\Gamma}(S_R^{[l]})_{l+1,\mathrm{div}}^{(1)})-
\int dx\Big(\theta\frac{\delta}{\delta\mathcal{B}}\Big)
\tilde{\Gamma}(S_R^{[l]})_{l+1,\mathrm{div}}^{(1)}=0, \\
\nonumber
&&2\xi\frac{\partial}{\partial\xi}\tilde{\Gamma}(S_R^{[l]})_
{l+1,\mathrm{div}}^{(1)}=(\tilde{\Gamma}_{0}^{(1)},\tilde{\Gamma}(S_R^{[l]})_
{l+1,\mathrm{div}}^{(2)})-(\tilde{\Gamma}_{0}^{(2)},\tilde{\Gamma}(S_R^{[l]})_
{l+1,\mathrm{div}}^{(1)})- \\
\label{ca11}
&&-\int dx\Big(\theta\frac{\delta}{\delta\mathcal{B}}\Big)\tilde{\Gamma}
(S_R^{[l]})_{l+1,\mathrm{div}}^{(2)}+\int dx\Big(\Omega_{m|}^{\ast}
\frac{\delta}{\delta\Omega_{m|}^{\ast}}-\Omega_{m|}\frac{\delta}
{\delta\Omega_{m|}}\Big)\tilde{\Gamma}(S_R^{[l]})_{l+1,\mathrm{div}}^{(1)}, \\
\label{ca12}
&&\tilde{\Gamma}(S_R^{[l]})_{l+1,\mathrm{div}}^{(k)}
\overleftarrow{h^\alpha}\omega_{\alpha}=0,  \\
\label{ca13}
&&\tilde{\Gamma}(S_R^{[l]})_{l+1,\mathrm{div}}^{(k)}\overleftarrow{T^\alpha}=0,
\quad k=1,2.
\eeq
Notice that the form of the equations  (\ref{ca10}) - (\ref{ca13}) does not depend on
the label $l$.

By taking into account the quantum numbers, axial-, Poincare-, $T$-symmetries,
the general expression for local functional
$\tilde{\Gamma}(S_R^{[l]})_{l+1,\mathrm{div}}^{(2)}$, reads
\beq
\label{ca14}
\tilde{\Gamma}(S_R^{[l]})_{l+1,\mathrm{div}}^{(2)}=\int
dx\left(q_{1,l+1}\mathcal{A}^{\ast}\mathcal{A}+q_{2,l+1}C^{\ast}C+q_{3,l+1}
\psi^{\ast}\psi+q_{4,l+1}\overline{\psi}^{\;\ast}\overline{\psi}+
q_{1,l+1}^{\prime}\mathcal{A}^{\ast}\mathcal{B}\right),
\eeq
where $q_{i,l+1}, i=1,2,3,4,$ и $q_{1,l+1}^{\prime}$ are arbitrary constants.
Then, by using the equation (\ref{ca12})
for $\tilde{\Gamma}(S_R^{[l]})_{l+1,\mathrm{div}}^{(2)}$, we find that $q_{1,l+1}^{\prime}=0$.
The final expression for
$\tilde{\Gamma}(S_R^{[l]})_{l+1,\mathrm{div}}^{(2)}$ has the form
\beq
\label{ca15}
\tilde{\Gamma}(S_R^{[l]})_{l+1,\mathrm{div}}^{(2)}=\int dx\left(q_{1,l+1}
\mathcal{A}^{\ast}\mathcal{A}+q_{2,l+1}C^{\ast}C+q_{3,l+1}\psi^{\ast}\psi+
q_{4,l+1}\overline{\psi}^{\;\ast}\overline{\psi}\right),
\eeq
Notice that the functional
$\tilde{\Gamma}(S_R^{[l]})_{l+1,\mathrm{div}}^{(2)}$ does not depend on the fields
$\theta$ and $\mathcal{B}$.

With the expression
(\ref{ca15}) for  $\tilde{\Gamma}(S_R^{[l]})_{l+1,
\mathrm{div}}^{(2)}$, the equation уравнение (\ref{ca11}) reduces to the
following one,
\beq
\nonumber
&&2\xi\frac{\partial}{\partial\xi}\tilde{\Gamma}(S_R^{[l]})_{l+1,
\mathrm{div}}^{(1)}-\int dx\left[q_{1,l+1}\left(\mathcal{A}\frac{\delta}
{\delta\mathcal{A}}-\mathcal{A}^{\ast}\frac{\delta}{\delta\mathcal{A}^{\ast}}
\right)+q_{2,l+1}\left(C\frac{\delta}{\delta C}-C^{\ast}\frac{\delta}
{\delta C^{\ast}}\right)+ \right. \\
\label{ca16}
&&\qquad +q_{3,l+1}\left(\psi\frac{\delta}{\delta\psi}-\psi^{\ast}\frac{\delta}
{\delta\psi^{\ast}}\right)+q_{4,l+1}\left.\left(\overline{\psi}\frac{\delta}
{\delta\overline{\psi}}-\overline{\psi}^{\;\ast}\frac{\delta}
{\delta\overline{\psi}^{\;\ast}}\right)\right]\tilde{\Gamma}_{0}^{(1)}=0.
\eeq

\subsubsection{Solution to equation (\ref{ca10})
for $\tilde{\Gamma}(S_R^{[l]})_{l+1,\mathrm{div}}^{(1)}$}

Consider a solution to the equation (\ref{ca10}) for the functional
$\tilde\Gamma^{(1)}_{l+1,\mathrm{div}}=\tilde{\Gamma}(S_R^{[l]})_{l+1,\mathrm{div}}^{(1)}$,
represented  in the form,
\beq
\nonumber
&&\tilde{\Gamma}_{l+1,\mathrm{div}}^{(1)}
=M_{\theta,l+1} +M_{\Omega^*,l+1}+M_{\psi,l+1}+M_{\mathcal{AB},l+1}, \\
\label{da1}
&&M_{\Omega^*,l+1}=M_{\mathcal{A}^*,l+1}+M_{C^*,l+1}+M_{\psi^*,l+1}+
M_{\overline{\psi}^{\;\ast},l+1}.
\eeq
With  this aim, we find first the general form of the functional
$\tilde\Gamma^{(1)}_{l+1,\mathrm{div}}$, using the locality, the quantum numbers,
axial-, Poincare -, $T$-symmetries and partially  the gauge symmetry in the external
field  ${\cal B}$. In fact, all required calculations do copy ones performed in Section 3
when constructing the general form of the functional  $\tilde{P}^{(1)}$ (see formulas
 (\ref{d2}) - (\ref{d16}) with the obvious replacements like
$\tilde{P}_{\theta}^{(1)}\to M_{\theta}$). Here, we reproduce the final results only.
The functional   $M_{\theta,l+1}$ has the form
\beq
\label{daa3}
M_{\theta,l+1}=q_{5,l+1}\int dx\mathcal{A}^{\ast\alpha}_
{\mu}(x)\theta^{\alpha}_{\mu}(x)=q_{5,l+1}\int dx\mathcal{A}^{\ast}\theta.
\eeq
For the functionals linear in antifields we find
\beq
\label{da5}
&&M_{\mathcal{A}^*,l+1}=\int dx\left[q_{6,l+1}\mathcal{A}^{\ast\alpha}_{\mu}
D^{\alpha\beta}_{\mu}(\mathcal{B})C^{\beta}+gq_{7,l+1\beta\gamma}^{\alpha}
\mathcal{A}^{\ast\alpha}_{\mu}\mathcal{A}^{\beta}_{\mu}C^{\gamma}\right] ,  \\
\label{da6}
&&M_{C^*,l+1}=\int dx\frac{g}{2}q_{8,l+1\beta\gamma}^{\alpha}
C^{\ast\alpha}C^{\beta}C^{\gamma},  \\
\label{da7}
&&M_{\psi^*,l+1}=-\int dx gq_{9,l+1jk}^{\alpha}\psi_j^{\ast}
\psi_{k}C^{\alpha},  \quad
M_{\overline{\psi}^*,l+1}=\int dx gq_{10,l+1kj}^{\alpha}
 \overline{\psi}_{j}^{\ast}\overline{\psi}_{k}C^{\alpha},
\eeq
where the constants  $"q"$ satisfy the equations (we omit the lower case  $l+1$)
\beq
\label{da9}
&&F_{\gamma\delta}^{\alpha}q_{7\delta\sigma}^{\beta}-q_{7\gamma\delta}
^{\beta}F_{\delta\sigma}^{\alpha}=f^{\alpha\beta\lambda}
q_{7\gamma\sigma}^{\lambda}, \\
\label{da10}
&&F_{\gamma\delta}^{\alpha}q_{8\delta\sigma}^{\beta}-q_{8\gamma\delta}^{\beta}
F_{\delta\sigma}^{\alpha}=f^{\alpha\beta\delta}q_{8\gamma\sigma}^{\delta}, \\
\label{da11} &&t_{jl}^{\alpha }q_{9lk}^{\beta}-q_{9jl}^{\beta}t_{lk}^{\alpha}=
f^{\alpha\beta\gamma}q_{9jk}^{\gamma}, \\
\label{da12}
&&t_{kl}^{\alpha}q_{10lj}^{\beta}-q_{10kl}^{\beta}t_{lj}^{\alpha}=
f^{\alpha\beta\gamma}q_{10kj}^{\gamma}.
\eeq
For the functionals  $M_{\psi,l+1}$ we obtain
\beq
\label{da16}
&&M_{\psi,l+1}=\int dx\left[iq_{11,l+1}\overline{\psi}\gamma^{\mu}D_{\psi\mu}
(\mathcal{B})\psi+igq_{12,l+1,jk}^{\alpha}\overline{\psi}_{j}\gamma^{\mu}
\mathcal{A}_{\mu}^{\alpha}\psi_k-mq_{13,l+1}\overline{\psi}\;\!\psi\right] ,\\
&&\qquad\qquad\qquad\qquad
t_{jr}^{\alpha}q_{12,l+1,rk}^{\beta}-q_{12,l+1,rl}^{\beta}t_{rk}^{\alpha}=
f^{\alpha\beta\gamma}q_{12,l+1,jk}^{\gamma}.
\label{da17}
\eeq

As the coefficient of the $\theta_\mu^\alpha$ term should be zero, it follows
that the equation
\beq
\nonumber
&&q_{5,l+1}\pa_{\mathcal{A}}(\Gamma_{0\mathcal{A}^*}+\Gamma_{0\psi}+\Gamma_
{0\mathcal{AB}})+\pa_{\mathcal{A}}(M_{\mathcal{A}^*,l+1}+M_{\psi,l+1}+
M_{\mathcal{AB},l+1})- \\
\label{daa28}
&&-\pa_{\mathcal{B}}(M_{\mathcal{A}^*,l+1}+M_{\psi,l+1}+
M_{\mathcal{AB},l+1})=0,
\eeq
holds. As the coefficient of the $\overline{\psi}\psi$ vertex
of the equation (\ref{daa28}) should be zero, it follows that
\beq
\label{daa29}
q_{12,l+1,jk}^\alpha=q_{12,l+1}t_{jk}^\alpha, \;
q_{12,l+1}=q_{11,l+1}-q_{5,l+1}.
\eeq
In turn, as the coefficient of the $\mathcal{A}^{\ast}C$ vertex of the equation (\ref{daa28})
should be zero, it follows that
\beq
\label{daa30}
q_{7,l+1,\beta\gamma}^\alpha=q_{7,l+1}f^{\alpha\beta\gamma}, \;
q_{7,l+1}=q_{6,l+1}-q_{5,l+1}.
\eeq
When  inserting  the expressions found for the
$q_{12,l+1,jk}^\alpha$ and $q_{7,l+1,\beta\gamma}^\alpha$
coefficients  into the equation (\ref{daa28}),
it reduces to the following one
\beq
\label{daa31}
q_{5,l+1}\pa_{\mathcal{A}_\mu^\alpha}\Gamma_{0\mathcal{A}\mathcal{B}}+
\pa_{\mathcal{A}_\mu^\alpha}M_{\mathcal{A}\mathcal{B},l+1}-
\pa_{\mathcal{B}_\mu^\alpha}M_{\mathcal{A}\mathcal{B},l+1}=0.
\eeq
The general solution to the equation (\ref{daa31}) reads
\beq
\label{daa32}
M_{\mathcal{A}\mathcal{B},l+1}=-q_{5,l+1}\mathcal{A}\pa_{\mathcal{A}}
\Gamma_{0\mathcal{A}\mathcal{B}}(V)+M_{1,l+1}(V), \;
V=\mathcal{A}+\mathcal{B},
\eeq
where $M_{1,l+1}(V)$  stands for  an arbitrary functional of $V$, at the moment.

In addition, insertion of the expression (\ref{daa30}) for the
$q_{7,l+1,\beta\gamma}^\alpha$ coefficient into the relation (\ref{da5}) yields
\beq\label{daa5}
M_{\mathcal{A}^*,l+1}=q_{6,l+1}\Gamma_{0\mathcal{A}^*}(V)-
q_{5,l+1}\mathcal{A}\pa_{\mathcal{A}}\Gamma_{0\mathcal{A}^*}(V),
\eeq
and insertion (\ref{daa29}) for the $q_{12,l+1,jk}^\alpha$ coefficient into
(\ref{da16}) gives the expression for  $M_{\psi,l+1}$,
\beq\label{daa5a}
M_{\psi,l+1}=q_{11,l+1}\Gamma_{0\psi|1}-q_{5,l+1}\mathcal{A}\pa_{\mathcal{A}}
\Gamma_{0\psi|1}+q_{13,l+1}\Gamma_{0\psi|2}.
\eeq

At $\theta=0$ the equation (\ref{ca10}) is reduced to the one
\beq
\label{dab26}
\int dx(\Gamma_{0\Omega^*}+\Gamma_{0\psi}+\Gamma_{0\mathcal{AB}})
\left(\overleftarrow{\pa}_{\Omega}\pa_{\Omega^*}-
\overleftarrow{\pa}_{\Omega^*}\pa_{\Omega}\right)
(M_{\Omega^*,l+1}+M_{\psi,l+1}+M_{\mathcal{AB},l+1})=0,
\eeq
which is not more than linear in antifields.

As  the coefficient of the $\mathcal{A}^*D(\mathcal{B})CC$ vertex of the equation (\ref{dab26})
should be zero, it follows that
\beq
\label{dab26a}
q_{8,l+1,\beta\gamma}^\alpha=q_{8,l+1}f^{\alpha\beta\gamma}, \quad
q_{8,l+1}=q_{7,l+1}=q_{6,l+1}-q_{5,l+1}.
\eeq

Next, we consider the equations which follow from  (\ref{dab26}) for zero-valued antifields.
They split into the two sets of equations. In the first set of equations,
\beq
\label{dab27}
\Gamma_{0\psi}\overleftarrow{\pa}_{\Omega}\pa_{\Omega^*}M_{\Omega^*,l+1}-
\Gamma_{0\Omega^*}\overleftarrow{\pa}_{\Omega^*}\pa_{\Omega}M_{\psi,l+1}=0,
\eeq
all vertexes contain the spinor fields. In the second ones,
\beq
\label{dab28}
\Gamma_{0\mathcal{AB}}\overleftarrow{\pa}_{\mathcal{A}}\pa_{\mathcal{A}^*}
M_{\mathcal{A}^*,l+1}-\Gamma_{0\mathcal{A}^*}\overleftarrow{\pa}_
{\mathcal{A}^*}\pa_{\mathcal{A}}M_{\mathcal{AB},l+1}=0,
\eeq
vertexes are constructed of the fields $\mathcal{A}$, $\mathcal{B}$ and
their coordinate-derivatives only.

As the coefficient of the $\overline{\psi}\psi\pa_\mu C$ vertex of the equation (\ref{dab27})
should be zero, it follows
\beq
\label{da29}
q_{9,l+1,jk}^\alpha=q_{9,l+1}t_{jk}^\alpha, \quad q_{9,l+1}=q_{6,l+1}-q_{5,l+1}.
\eeq
As the coefficient of the $\overline{\psi}\psi C$ vertex of the equation (\ref{dab27})
should be zero,  it follows the relation,
\beq
\label{dab30}
q_{10,l+1,jk}^\alpha=q_{9,l+1,jk}^\alpha=q_{9,l+1}t_{jk}^\alpha.
\eeq

When inserting  the expressions (\ref{daa32}) and (\ref{daa5}) for
$M_{\mathcal{AB},l+1}$ and $M_{\mathcal{A}^*,l+1}$, respectively, into the equation
(\ref{dab28}), it reduces to the following equation,
\beq
\label{dab31}
D(V)\pa_V M_{1,j+1}(V)=0 \; \Rightarrow \;
M_{1,j+1}(V)=q_{14,l+1}\Gamma_{0\mathcal{AB}}.
\eeq

Thus, the functional $\Gamma(S_R^{[l]})_{l+1,\mathrm{div}}$ describing the  описывающий
$(l+1)$-loop divergences of the functional $\Gamma(S_R^{[l]})$, has the form
\beq
\Gamma(S_R^{[l]})_{l+1,\mathrm{div}}=\tilde{\Gamma}^{(1)}(S_R^{[l]})_{l+1,
\mathrm{div}}+\chi\tilde{\Gamma}^{(2)}(S_R^{[l]})_{l+1,\mathrm{div}},
\eeq
where the functional $\tilde{\Gamma}^{(2)}(S_R^{[l]})_{l+1,\mathrm{div}}$ is given by the
expression (\ref{ca15}). As for the functional $\tilde{\Gamma}^{(1)}(S_R^{[l]})_{l+1,
\mathrm{div}}$ we use the representation
\beq
\label{dab31a}
&&\tilde{\Gamma}(S_R^{[l]})_{l+1,\mathrm{div}}^{(1)}
=M_{\theta,l+1} +M_{\Omega^*,l+1}+M_{\psi,l+1}+M_{\mathcal{AB},l+1},
\eeq
where the functionals $"M"$,
\beq
\label{dab32a}
&&M_{\theta,l+1}=q_{5,l+1}\Gamma_{0\theta}, \; 
\\
&&M_{\mathcal{A}^*,l+1}=q_{6,l+1}\Gamma_{0\mathcal{A}^*}(V)-q_{5,l+1}
\mathcal{A}\pa_{\mathcal{A}}\Gamma_{0\mathcal{A}^*}(V),\;
 \\
&&M_{C^*,l+1}=(q_{6,l+1}-q_{5,l+1})\Gamma_{0C^*}, \;
\\
&&M_{\psi^*,l+1}=(q_{6,l+1}-q_{5,l+1})\Gamma_{0\psi^*}, \;
\\
&&M_{\overline{\psi}^*,l+1}=(q_{6,l+1}-q_{5,l+1})\Gamma_{0\overline{\psi}^*},\;
\\
&&M_{\psi,l+1}=q_{11,l+1}\Gamma_{0\psi|1}-q_{5,l+1}\mathcal{A}\pa_{\mathcal{A}}
\Gamma_{0\psi|1}+q_{13,l+1}\Gamma_{0\psi|2}, \;
\\
\label{dab32g}
&&M_{\mathcal{AB},l+1}=q_{14,l+1}\Gamma_{0\mathcal{AB}}-
q_{5,l+1}\mathcal{A}\pa_{\mathcal{A}}\Gamma_{0\mathcal{AB}}, \;
\eeq
are represented in terms of the tree loop functionals
$"\Gamma_0"$ (\ref{da19}) - (\ref{da25}).

\subsubsection{Solution to equation (\ref{ca16})
for $\tilde{\Gamma}(S_R^{[l]})_{l+1,\mathrm{div}}^{(1)}$}

When inserting the representation for the functional
$\tilde{\Gamma}^{(1)}(S_R^{[l]})_{l+1,\mathrm{div}}$ given by
(\ref{dab31a}) - (\ref{dab32g}) into the equations (\ref{ca16}), it
takes the form of zero value for some linear combinations of
structures appeared in the right-hand side of formulas
(\ref{dab32a}) - (\ref{dab32g}).

As the coefficient of the
$\theta_\mu^\alpha$ term should be zero, it follows that
\beq \label{e1}
q_{1,l+1}=-2\xi\dot{q}_{5,l+1}. \eeq

Then, as the  coefficient of  any
antifield term should be zero, it follows
\beq
\label{e2}
q_{2,l+1}=2\xi(\dot{q}_{6,l+1}-\dot{q}_{5,l+1}).
\eeq

When the relations (\ref{e1}) and (\ref{e2}) hold, then the equation  (\ref{ca16})
reduces to the two equations having obvious solutions
\beq
\label{e3}
2\xi\dot{q}_{14,l+1}\Gamma_{0\mathcal{AB}}=0 \; \Rightarrow \;
\dot{q}_{14,l+1}=0,
\eeq
\beq
&&2\xi\dot{q}_{11,l+1}\Gamma_{0\psi|1}+2\xi\dot{q}_{11,l+1}\Gamma_{0\psi|2}-
(q_{3,l+1}+q_{4,l+1})\left(\Gamma_{0\psi|1}+\Gamma_{0\psi|2}\right)=0 \;
\Rightarrow \\
\label{e4}
&&\qquad\qquad\qquad 2\xi\dot{q}_{11,l+1}=2\xi\dot{q}_{13,l+1}=q_{3,l+1}+q_{4,l+1}.
\eeq

It is convenient to introduce new parameters
$q_{15,l+1}$ and $q_{16,l+1}$,
\beq
q_{15,l+1}=q_{13,l+1}-q_{11,l+1}, \quad 2q_{16,l+1}=q_{3,l+1}-q_{4,l+1},
\eeq
in terms of which the equations  (\ref{e4}) rewrites as
\beq
\label{e5}
\dot{q}_{15,l+1}=0, \quad q_{3,l+1}=\xi\dot{q}_{11,l+1}+q_{16,l+1}, \quad
q_{4,l+1}=\xi\dot{q}_{11,l+1}-q_{16,l+1}.
\eeq

\subsection{Finiteness  of $\Gamma$ to (l+1)-loop approximation}

Now,let us prove that one can chose the renormalization constants in  such a way as to make
 the effective action finite to the $(l+1)$-loop approximation.
 To this end, we consider the divergent
part of the effective action $\Gamma$, $\Gamma_{l+1,\mathrm{div}}$, described
by the equation (\ref{ca9}),
\beq
\label{e6}
\Gamma_{l+1,\mathrm{div}}=\Gamma(S_R^{[l]})_{l+1,\mathrm{div}}+s_{l+1}=
\Gamma_{l+1,\mathrm{div}}^{(1)}+\chi\Gamma_{l+1,\mathrm{div}}^{(2)}.
\eeq
For the functionals $\Gamma_{l+1,\mathrm{div}}^{(1)},\;\Gamma_{l+1,\mathrm{div}}^{(2)}$
we have the representations
\beq
\nonumber
&&\Gamma_{l+1,\mathrm{div}}^{(1)}=\Gamma(S_R^{[l]})_{l+1,\mathrm{div}}^{(1)}+
s_{l+1}^{(1)}=\\
\label{e8}
&&\qquad\quad\;\; =\Gamma_{\theta,l+1,\mathrm{div}}+
\Gamma_{\Omega^*,l+1,\mathrm{div}}+\Gamma_{\psi,l+1,\mathrm{div}}+
\Gamma_{\mathcal{AB},l+1,\mathrm{div}}, \\
\label{e9a}
&&\Gamma_{\theta,l+1,\mathrm{div}}=(q_{5,l+1}+z_{5,l+1})\Gamma_{0\theta}, \\
&&\Gamma_{\mathcal{A}^*\theta,l+1,\mathrm{div}}=(q_{6,l+1}+z_{6,l+1})
\Gamma_{0\mathcal{A}^*}(V)-(q_{5,l+1}+z_{5,l+1})\mathcal{A}\pa_{\mathcal{A}}
\Gamma_{0\mathcal{A}^*}(V), \\
&&\Gamma_{C^*,l+1,\mathrm{div}}=(q_{6,l+1}+z_{6,l+1}-q_{5,l+1}-z_{5,l+1})
\Gamma_{0C^*}, \\
&&\Gamma_{\psi^*,l+1,\mathrm{div}}=q_{6,l+1}+z_{6,l+1}-q_{5,l+1}-z_{5,l+1})\Gamma_{0\psi^*}, \\
&&\Gamma_{\overline{\psi}^*,l+1,\mathrm{div}}=q_{6,l+1}+z_{6,l+1}-q_{5,l+1}-
z_{5,l+1})\Gamma_{0\overline{\psi}^*}, \\
\nonumber
&&\Gamma_{\psi,l+1,\mathrm{div}}=(q_{11,l+1}+z_{11,l+1})\Gamma_{0\psi|1}-
(q_{5,l+1}+z_{5,l+1})\mathcal{A}\pa_{\mathcal{A}}\Gamma_{0\psi|1}+ \\
&&\qquad\qquad\quad+(q_{11,l+1}+z_{11,l+1}+q_{15,l+1}+z_{15,l+1})\Gamma_{0\psi|2}, \\
\label{e9i}
&&\Gamma_{\mathcal{AB},l+1,\mathrm{div}}=(q_{14,l+1}+z_{14,l+1})
\Gamma_{0\mathcal{AB}}-(q_{5,l+1}+z_{5,l+1})\mathcal{A}\pa_{\mathcal{A}}
\Gamma_{0\mathcal{AB}},
\eeq
\beq
\nonumber
&&\Gamma_{l+1,\mathrm{div}}^{(2)}=\Gamma(S_R^{[l]})_{l+1,\mathrm{div}}^{(2)}+
s_{l+1}^{(2)}=\int dx\Big[2\xi(\dot{q}_{5,l+1}+\dot{z}_{5,l+1})\mathcal{A}^
{\ast}\mathcal{A}+ \\
\nonumber
&&+2\xi(\dot{q}_{6,l+1}+\dot{z}_{6,l+1}-\dot{q}_{5,l+1}-\dot{z}_{5,l+1})
C^{\ast}C+(\xi\dot{q}_{11,l+1}+\xi\dot{z}_{11,l+1}+q_{16,l+1}+ \\
\label{e7}
&&+z_{16,l+1})\psi^{\ast}\psi+(\xi\dot{q}_{11,l+1}+\xi\dot{z}_{11,l+1}-
q_{16,l+1}-z_{16,l+1})\overline{\psi}^{\;\ast}\overline{\psi}\Big].
\eeq

It follows from the formulas (\ref{e6}) - (\ref{e7}), that the choice of the parameters
 $z_{i,l+1}$ in the form
форме
\beq
\label{e10}
z_{i,l+1}=z_{i,l+1,\mathrm{fin}}, \quad z_{i,l+1,\mathrm{fin}}=-q_{i,l+1}, \quad
i=5,6,11,14,15,16,
\eeq
provides for zero-valued coefficients to the  $(l+1)$-loop  divergences,
\beq
\label{e11}
\Gamma_{l+1,\mathrm{div}}\Big|_{z_{i,l+1}=z_{i,l+1,\mathrm{fin}}}=0, \quad
\Gamma_{l+1}\Big|_{z_{i,l+1}=z_{i,l+1,\mathrm{fin}}}=
\Gamma_{l+1,\mathrm{fin}}.
\eeq

Notice that the choice of parameters $z_{i,l+1,\mathrm{fin}}$ is unique within the
minimal subtraction scheme.

\subsection{(l+2)-loop approximation}

The renormalization of  $S_R$ to the  $(l+1)$-loop approximation allows one
to construct the effective action
$\Gamma$, finite to that approximation; however it does not satisfy exactly
the extended master-equation and the gauge dependence equation, by itself.
We show the possibility to complete the renormalization constants of the action $S_R$
with the help of the  $(l+2)$-loop approximation, so that  it will satisfy the equations
mentioned to the $(l+1)$-loop approximation and, in its turn, the corresponding
effective action, finite to the $(l+1)$-loop approximation, will satisfy the set of equations
(\ref{a21}) - (\ref{a25}) to that approximation.

Indeed, we represent the action $S_R$ as
\beq
S_R=S_R^{[l+1]}+\eta^{l+2}s_{l+2}+O(\eta^{l+3}),
\eeq
where  $S_R^{[l+1]}$ is the action$S_R$ with independent parameters с $Z_i$
replaced by $Z_i^{[l+1]}$, and $s_{l+2}$ is equal to
\beq
s_{l+2}=s_{l+2}^{(1)}+\chi s_{l+2}^{(2)},
\eeq
where
\beq
&&s_{l+2}^{(1)}=s_{\theta,l+2}+s_{\Omega^*,l+2}+s_{\psi,l+2}+
s_{\mathcal{AB},l+2}, \\
&&s_{\theta,l+2}=z_{5,l+2}\Gamma_{0\theta}, \\
&&s_{\mathcal{A}^*,l+2}=z_{6,l+2}\Gamma_{0\mathcal{A}^*}-
z_{5,l+2}\mathcal{A}\pa_{\mathcal{A}}\Gamma_{0\mathcal{A}^*}, \\
&&s_{C^*,l+2}=(z_{6,l+1}-z_{5,l+2})\Gamma_{0C^*}, \\
&&s_{\psi^*,l+2}=(z_{6,l+2}-z_{5,l+2})\Gamma_{0\psi^*}, \;
s_{\overline{\psi}^{\;\ast},l+2}=(z_{6,l+2}-z_{5,l+2})
\Gamma_{0\overline{\psi}^{\;\ast}}, \\
&&s_{\psi,l+2}=z_{11,l+2}\Gamma_{0\psi|1}-z_{5,l+2}\mathcal{A}
\pa_{\mathcal{A}}\Gamma_{0\psi|1}+(z_{11,l+2}+z_{15,l+2})\Gamma_{0\psi|2}, \\
&&s_{\mathcal{AB},l+2}=z_{14,l+2}\Gamma_{0\mathcal{AB}}-
z_{5,l+2}\mathcal{A}\pa_{\mathcal{A}}\Gamma_{0\mathcal{AB}},
\eeq
\beq
\nonumber
&&s_{l+2}^{(2)}=\int dx\Big[2\xi\dot{z}_{5,l+2}\mathcal{A}^{\ast}\mathcal{A}+
2\xi(\dot{z}_{6,l+2}-\dot{z}_{5,l+2})C^{\ast}C+\\
&&\qquad\qquad+\xi(\dot{z}_{11,l+2}+
z_{16,l+2})\psi^{\ast}\psi+ +\xi(\dot{z}_{11,l+2}-z_{16,l+2})\overline{\psi}^{\;\ast}
\overline{\psi}\Big],\\
&&{\dot z}_{14,l+2}={\dot z}_{15,l+2}=0.
\eeq

Notice that the action $S_R^{[l+1]}$ satisfies the equations
(\ref{a21}) - (\ref{a25}).

Further calculations and consequences from them do copy exactly
the results of the previous subsection
with the natural replacement $l+1 \; \rightarrow \; l+2$.

Also, it is obvious that the procedure of divergence
compensations discussed can be applied to the case  $l=0$ so that by using the loop induction
method in Feynman diagrams for the functional $\Gamma$, we arrive at the following statement:
for the $l$-loop approximation   $\Gamma^{[l]}$, where
$l$ is arbitrary positive integer,
\beq
\Gamma^{[l]}=\sum_{n=0}^l\eta^n\Gamma_n,
\eeq
of the functional  $\Gamma$ defined by the relations (\ref{b1}), (\ref{b2}),
there exists the uniquely  defined parameters $Z_i^{[l]}$, $i=5,6,11,14,15,16,$
\beq
\label{e12}
\dot{Z}_{14}^{[l]}=0, \quad \dot{Z}_{15}^{[l]}=0, \; \forall l\geq0,
\eeq
such that the functional $\Gamma^{[l]}$ does not contain divergences
and $\Gamma$ satisfies the equations (\ref{b4}) - (\ref{b7}).

\section{Relations between parameters of $S_R$ and standard renormalization constants}

In that section we find relations between some parameters of the action $S_R$
and the standard renormalization constants.
Within  the expression for $S_R$, we restrict ourselves only by desired vertexes in symbolic
notation
\beq
\label{s1} S_R=\int
dx\left(Z_{14}Z_5^{-2}\pa A\pa A+gZ_{14}Z_5^{-3}A^2\pa A+
Z_{11}\overline{\psi}\pa\psi+mZ_{13}\overline{\psi}\psi+...\right),
\eeq
where the ellipsis means the rest vertexes. As the propagators  of fields
 $A$ and $\psi$ are finite, they should be considered as renormalized fields.
Then, we find:
\beq
Z_A=Z_{14}^{1/2}Z_5^{-1}, \quad Z_\psi=Z_{11}^{1/2},
\eeq
where $Z_A$ and $Z_\psi$ are the renormalization constants of the bare fields
$A_0$ and $\psi_0$. The coefficient of the second vertex
in the expression (\ref{s1}) gives the renormalization for vertex $A^3$,
\beq Z_{A^3}=Z_{14}Z_5^{-3} \;
\Rightarrow \; g_0=Z_gg, \; Z_g=Z_{A^3}Z_A^{-3}= Z_{14}^{-1/2}.
\eeq
The coefficient of the forth vertex
in the expression (\ref{s1}) gives the renormalization for vertex $\overline{\psi}\psi$,
\beq
Z_{\overline{\psi}\psi}=Z_{13} \; \Rightarrow \; m_0=Z_mm, \;
Z_m=Z_{\overline{\psi}\psi}Z_\psi^{-2}=Z_{13}/Z_{11}=Z_{15}.
\eeq
It follows from the equations (\ref{e12}) that the renormalization constants
of physical parameters  $g$ and $m$ do not depend on gauge,
\beq
\pa_\xi Z_g=0, \quad \pa_\xi Z_m=0.
\eeq

\section{Summary}
\noindent
In the present paper, within the background field formalism,
it is studied the renormalization procedure and the gauge
dependence of the theory of Yang-mills fields interacting
with a multiplet of massive spinor fields. It is shown that
the extension of the Faddeev-Popov action with extra fields
and parameters allows one to establish the multiplicative
character of the renormalizability. The proofs given above
are based on the possibility to expand the effective action
in loops, as well as to use the minimal subtraction scheme
as to eliminate divergences. It is a new and important result
that the renormalization constant of the  mass parameter
is shown to be gauge-independent.

\section*{Acknowledgments}
\noindent  The work of I.A. Batalin and I.V. Tyutin is
supported in part by the RFBR grant 17-02-00317. The work of P.M.
Lavrov is supported partially by the Ministry of Education and Science of
the Russian Federation, grant  3.1386.2017 and by the RFBR grant
18-02-00153.
\\

\end{document}